%% file: HD4539_mnras.tex
\documentclass[fleqn,usenatbib]{mnras}
\usepackage[T1]{fontenc}
\usepackage{ae,aecompl}
\usepackage{hyperref}

\usepackage{graphicx}	% Including figure files
\usepackage{amsmath}	% Advanced maths commands
\usepackage{amssymb}	% Extra maths symbols

\input{defs}  % newcommands in defs.tex

\title[High-degree g-modes in the single sdB star HD\,4539]
{High-degree gravity modes in the single sdB star HD\,4539}
\author[]{
R. Silvotti,$^{1}$\thanks{E-mail: roberto.silvotti@inaf.it}
M. Uzundag,$^{1,2}$
A.\,S. Baran,$^{3}$
R.\,H. \O stensen,$^{4}$
J.\,H. Telting,$^{5}$
\newauthor
\hspace{0.0mm}
U. Heber,$^{6}$
M.\,D. Reed,$^{4}$
M. V\v{u}ckovi\'{c}$^{2}$
\vspace{2mm}
\\
$^{1}$INAF-Osservatorio Astrofisico di Torino, Strada dell'Osservatorio 20, 
10025, Pino Torinese, Italy\\
$^{2}$Instituto de Fisica y Astronomia, Facultad de Ciencias, Universidad de 
Valparaiso, Gran Bretana 1111, Playa Ancha, 2360102,\\ 
\hspace{0.0mm}
Valparaiso, Chile\\
$^{3}$Uniwersytet Pedagogiczny, Obserwatorium na Suhorze, ul. 
Podchor\c{a}\.zych 2, 30-084 Krak\'ow, Polska\\
$^{4}$Department of Physics, Astronomy and Materials Science, Missouri State 
University, 901 S. National, Springfield, MO 65897, USA\\
$^{5}$Nordic Optical Telescope, Rambla José Ana Fernández Pérez 7, 
E-38711 Breña Baja, Spain\\
$^{6}$Dr. Remeis-Sternwarte \& ECAP, Astronomical Institute, University of 
Erlangen-N\"{u}rnberg, Sternwartstr. 7, D-96049 Bamberg,\\ 
\hspace{0.0mm}
Germany
}

\date{Accepted XXX. Received YYY; in original form ZZZ}

\pubyear{2019}

\begin{document}
\label{firstpage}
\pagerange{\pageref{firstpage}--\pageref{lastpage}}
\maketitle

% Abstract of the paper
\begin{abstract}
HD\,4539 (alias PG\,0044+097 or EPIC\,220641886) is a bright (V=10.2) 
long-period V1093~Her-type subdwarf~B (sdB) pulsating star that was observed 
by the \kepler\ spacecraft in its secondary ($K2$) mission.
We use the $K2$ light curve (78.7~days) to extract 169 pulsation 
frequencies, 124 with a robust detection.
Most of these frequencies are found in the low-frequency region typical of 
gravity (g-)modes, but some higher frequencies corresponding to 
pressure (p-)modes are also detected.
Therefore HD\,4539 is a hybrid pulsator and both the deep and surface layers 
of the star can potentially be probed through asteroseismology.
The lack of any frequency splitting in its amplitude spectrum
suggests that HD\,4539 has a rotation period longer than the $K2$ run and/or 
that it is seen pole-on.
From asymptotic period spacing we see many high-degree modes, up to $l$=12, 
in the spectrum of HD\,4539, with amplitudes as low as a few ppm.
A large fraction of these modes 
can be identified and for $\sim$29\% of them we obtain a unique and 
robust identification corresponding to $l$$\le$8.
Our study includes also a new determination of the atmospheric parameters
of the star.
From low-resolution spectroscopy we obtain 
\teff=22,800$\pm$160~K, \logg=5.20$\pm$0.02 and 
log($N$(He)/$N$(H))=--2.34$\pm$0.05.
By fitting the SED we obtain 
\teff=23,470$^{+650}_{-210}$~K, R$_\star$=0.26$\pm$0.01~\rsun\ and 
M$_\star$=0.40$\pm$0.08~\msun.
Moreover, from 11 high-resolution spectra we see the radial velocity variations
caused by the stellar pulsations, 
with amplitudes of $\approx$150~m/s for the main modes, and we can exclude
the presence of a companion with a minimum mass higher than a few Jupiter 
masses for orbital periods below $\sim$300~days.
\end{abstract}

% Select between one and six entries from the list of approved keywords.
% Don't make up new ones.
\begin{keywords}
stars: horizontal branch; stars: oscillations (including pulsations);\\
asteroseismology.
\end{keywords}

%%%%%%%%%%%%%%%%%%%%%%%%%%%%%%%%%%%%%%%%%%%%%%%%%%

%%%%%%%%%%%%%%%%% BODY OF PAPER %%%%%%%%%%%%%%%%%%

\section{Introduction}

Hot subdwarf stars of spectral class B (sdB) are core helium-burning stars, 
found both in the disk and halo of our Galaxy. 
Their observed properties locate them in the extreme horizontal branch (EHB) 
part of the H-R diagram, with effective temperatures from $\sim$22,000 to 
$\sim$38,000~K and surface gravities of 5.0\,\lsim\,\logg\,\lsim\,6.2 in cgs 
units. 
They are compact objects with radii on the order of 0.2 $R_{\odot}$ and 
typical masses around 0.47\,$M_{\odot}$. 
These stars have experienced extreme mass loss near the tip of the red giant 
branch, when nearly the entire hydrogen envelope was lost, leaving a 
helium-burning core with a very thin inert hydrogen-rich envelope 
($M_{env}$\,\lsim\,0.01\,$M_{\odot}$), too thin to sustain hydrogen shell 
burning and ascend the asymptotic giant branch.
The mechanisms responsible for this extreme mass loss are not yet well 
understood but it is clear that binarity plays a major role, at least for 
half sdB stars, those in a close binary with a white dwarf or an M-dwarf
companion. The merger of two helium white dwarfs is another option to form
single hot subdwarfs but only a small fraction of single sdB stars have masses
large enough to be compatible with this mechanism (see e.g. discussion and 
Fig.~6 of \citealt{fontaine12}).
Thus the formation of single sdB stars is still an open question.
After depletion of helium in the core, sdB stars may evolve into subdwarf O 
(sdO) stars burning helium in a shell surrounding the C/O core (although a 
direct evolutionary link between sdB and sdO stars remains uncertain), and 
eventually they will end their evolutionary journey directly as a white dwarf
(see \citealt{heber16} for a review).

About 10\% of sdB stars with effective temperatures (\teff) between 
$\sim$28,000 and $\sim$36,000~K are found to pulsate with short period 
p-modes of a few minutes \citep{ostensen10}.
They are termed V361\,Hya stars from the prototype \citep{kilkenny97}, that 
was discovered at about the same time at which p-modes were 
predicted by theory \citep{charpinet96}.
At temperatures below 28,000~K, about 75\%\ of sdB stars are found to be 
pulsating \citep{ostensen11} with longer-period g-modes, from $\sim$45 min up 
to few hours, and they are termed V1093\,Her stars from the prototype 
\citep{green03}.
Both p- and g-mode pulsations are driven by cyclic ionization of iron-group 
elements \citep{charpinet97, fontaine03}, which are pushed up by radiative 
levitation.
Near the boundary between the two instability strips, at \teff $\approx$ 
28,000~K, the so-called hybrid pulsators, that show both p- and g-modes 
\citep{baran05, schuh06}, offer the opportunity to probe both the external 
layers and the core of sdB stars.

A substantial improvement in our understanding of the stellar oscillations 
in sdB stars was brought by the \kepler\ space telescope \citep{borucki10, 
gilliland10}, that observed continuously 18 sdB pulsators for many months (up 
to more than 3 years) in its primary mission, and more than 30 pulsators for 
70/90 days in its secondary mission ($K2$, \citealt{howell14}), after the 
second reaction wheel failed. 
The sampling time of 58~s of the so-called short-cadence data was not ideal 
for the short-period p-modes but very good for the longer-period g-modes.
Thanks to the high-quality data of \kepler/$K2$, the constant period spacing 
predicted for the g-modes in the asymptotic limit was observed in 
several pulsators, as well as the missing periods due to trapped modes 
(\citealt{reed11}; \citealt{kern18}; \citealt{reed18}, and references 
therein).
Multiplets of evenly spaced frequencies have confirmed that sdB stars are
slow rotators, with typical periods of several days or tens of days, 
up to 100 days or more \citep{reed18, charpinet18}.
Through period and frequency spacing it was possible to identify a large
fraction of the pulsation modes detected in several stars and high-degree 
g-modes were seen in a few objects, up to $l$=6 \citep{kern18}
or $l$=8 \citep{telting14a}.
The huge quantity of novel information contained in the \kepler/$K2$ data 
represents a strong challenge for seismic models and has also opened new
possibilities of studying the evolution of the amplitude spectra in time,
with variations of amplitudes and frequencies that are likely related
to nonlinear interactions between different pulsation modes 
(see e.g. \citealt{zong18}).
Since July 2018, the NASA $TESS$ ({\it Transiting Exoplanet Survey 
Satellite}) mission is continuing the work started by \kepler/$K2$ observing 
many new sdB pulsators and the first results are published in
\citet{charpinet19}.

In this article we describe the results of an analysis of both
photometric and spectroscopic data of the sdB pulsator HD\,4539 
(alias PG\,0044+097 or EPIC\,220641886).
Pulsations in this star were previously detected by \citet{schoenaers07} 
from line profile variations of 198 spectra acquired with the Grating 
Spectrograph at the 1.9~m SAAO telescope.
The $K2$ light curve confirms the presence of pulsations and shows a very 
rich spectrum with many high-degree low-amplitude g-modes and a few p-modes, 
making this bright sdB star (V=10.2) one of the most interesting objects 
for asteroseismic studies.

%\newpage

\section{K2 photometry} 

\begin{figure}
\centering
\includegraphics[width=8.6cm,angle=0]{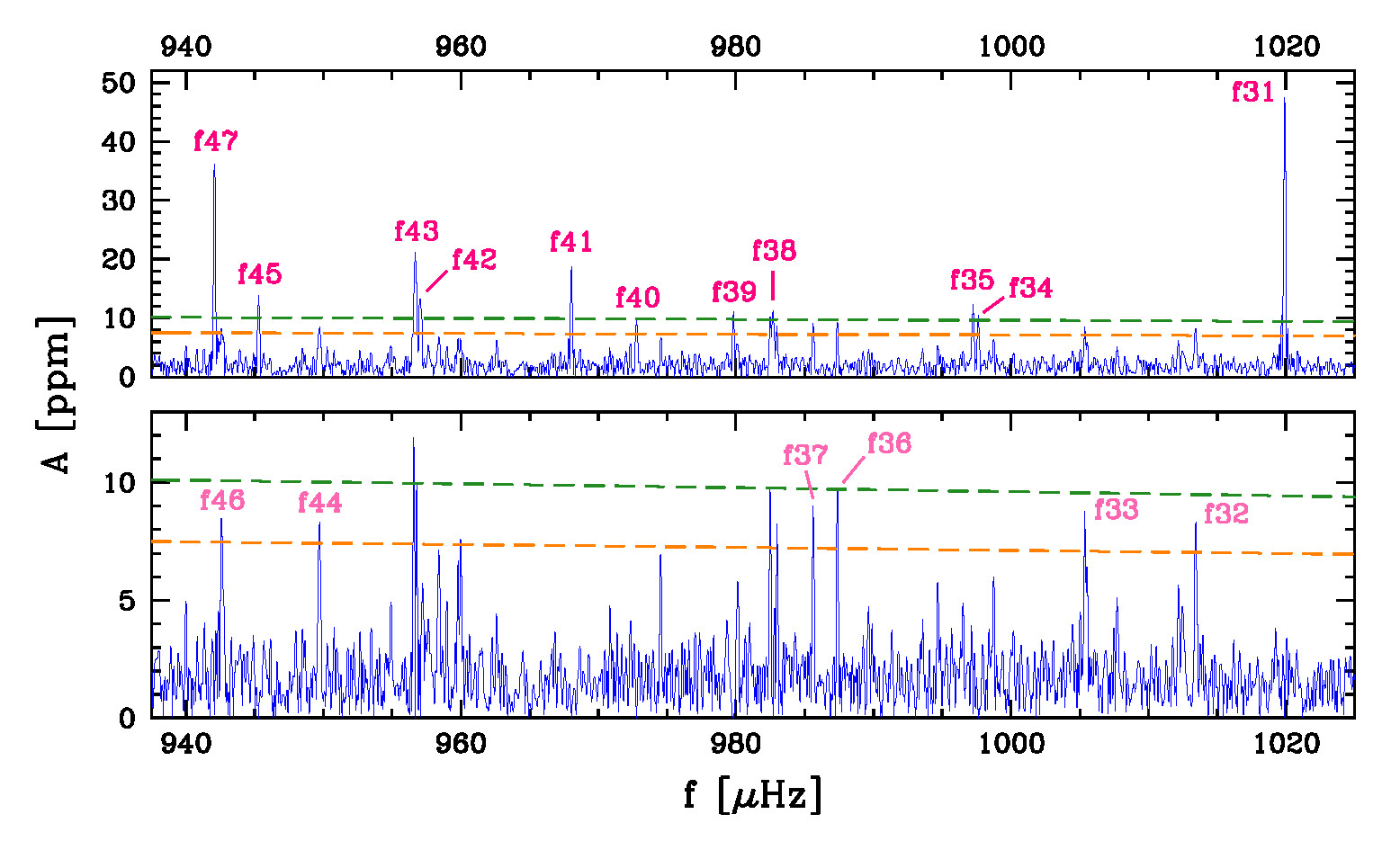}
\vspace{-4mm}
\caption{The 2-step procedure that we used to extract the pulsation 
frequencies.
Top panel: the main frequencies (those with an amplitude higher than 
5.4\,$\sigma$, 124 in total) are extracted from the amplitude spectrum of the 
data.
Bottom panel: once the 124 main frequencies have been subtracted from the 
data, secondary frequencies (45 in total, those with an amplitude between 4 
and 5.4\,$\sigma$) are extracted from the amplitude spectrum of the residuals.
The thresholds of 4 and 5.4\,$\sigma$ are represented by the orange and green 
lines respectively. 
Note that the peaks near 957 and 983~$\mu$Hz are not considered since they
are just the residuals of f43 and f38, due to temporal variations of the
frequencies and/or unresolved close frequencies.
Instead, the peak at 960~$\mu$Hz is not considered since it was below
4\,$\sigma$ in the upper panel.
More details in the text.}
\label{fig1}
\end{figure}

\begin{table*} 
\centering
\caption[]{Pulsation frequencies.}
\begin{tabular}{lrrrrll}
\hline
{\bf ID} & \multicolumn{1}{c}{\bf F} & \multicolumn{1}{c}{\bf P} & \multicolumn{1}{c}{\bf A}
   & \multicolumn{1}{c}{\bf l} & \multicolumn{1}{c}{\bf n} & {\bf Notes}\\
   & \multicolumn{1}{c}{[\muHz]} & \multicolumn{1}{c}{[s]} 
   & \multicolumn{1}{c}{[ppm]} &    &	  & \\
\hline
                 f1   & 4531.833 &   220.66 &  13.1~ & --- & --- & \\
\hspace{-2.2mm} (f2   & 3503.332 &   285.44 &   6.2~ & --- & ---)& \\
\hspace{-2.2mm} (f3   & 3499.973 &   285.72 &	5.2~ & --- & ---)& \\ 
                 f4   & 3497.986 &   285.88 &	7.3~ & --- & --- & MR \\
\hspace{-2.2mm} (f5   & 3133.521 &   319.13 &	5.2~ & --- & ---)& \\ 
                 f6   & 2969.824 &   336.72 &	8.6~ & --- & --- & \\
                 f7   & 2969.661 &   336.74 &	7.6~ & --- & --- & \\
                 f8   & 2968.864 &   336.83 &	9.8~ & --- & --- & \\
                 f9   & 2968.701 &   336.85 &	7.3~ & --- & --- & \\
                 f10  & 2968.444 &   336.88 &  12.4~ & --- & --- & \\
\hspace{-2.2mm} (f11  & 2963.608 &   337.43 &	5.3~ & --- & ---)& \\ 
                 f12  & 2945.414 &   339.51 &	7.5~ & --- & --- & \\
\hspace{-2.2mm} (f13  & 2941.457 &   339.97 &	6.0~ & --- & ---)& \\ 
\hspace{-2.2mm} (f14  & 2516.795 &   397.33 &	4.9~ & --- & ---)& \\ 
                 f15  & 2329.916 &   429.20 &	9.4~ & --- & --- & MR \\
\multicolumn{7}{l}{\bf \hspace{-2.2mm} ------~ approximate boundary between p- and g-modes ~------}\\
\hspace{-2.2mm} (f16  & 1704.969 &   586.52 &	6.3~ & --- & ---)& \\
\hspace{-2.2mm} (f17  & 1683.290 &   594.07 &	5.5~ & --- & ---)& \\ % [l=9 n=17?]
                 f18  & 1465.086 &   682.55 &	7.7~ & --- & --- & \\ % [l=12 n=26? or l=6 n=13?]
                 f19  & 1397.473 &   715.58 &  10.4~ & --- & --- & \\ % [l=7 n=16?]
                 f20  & 1246.915 &   801.98 &	9.0~ & --- & --- & \\ % [l=10 n=25?]
                 f21  & 1227.676 &   814.55 &	8.3~ & --- & --- & \\
                 f22  & 1227.170 &   814.88 &	9.5~ & --- & --- & \\
\hspace{-2.2mm} (f23  & 1226.799 &   815.13 &	6.0~ & --- & ---)& \\ % [l=12 n=31? or l=9 n=23?]
                 f24  & 1166.801 &   857.04 &	7.6~ & --- & --- & MR \\ % [l=9 n=24? or l=7 n=19?]
\hspace{-2.2mm} (f25  & 1105.843 &   904.29 &	6.2~ &  10 &  28)& TI \\
\hspace{-2.2mm} (f26  & 1077.765 &   927.85 &	6.6~ &   9 &  26)& or l=12, n=35 \\
\hspace{-2.2mm} (f27  &	1067.695 &   936.60 &	7.0~ & --- & ---)& \\ % NEW !!!!!!!!!!!
                 f28  & 1067.503 &   936.77 &	9.7~ &  10 &  29 & TI \\
                 f29  & 1034.893 &   966.28 &  22.8~ &   9 &  27 & TI; NoR \\
                 f30  & 1025.764 &   974.88 &  11.6~ & --- & --- & \\ % [l=10, n=30]
                 f31  & 1019.899 &   980.49 &  48.4~ &  12 &  37 & TI; NoR \\
\hspace{-2.2mm} (f32  & 1013.432 &   986.75 &	8.5~ & --- & ---)& \\
\hspace{-2.2mm} (f33  & 1005.356 &   994.67 &	8.7~ &   7 &  22)& or l=8, n=24 \\
                 f34  &  997.620 &  1002.39 &	9.7~ &   9 &  28 & or l=10, n=31 \\
                 f35  &  997.230 &  1002.78 &  12.5~ &  10 &  31 & or l=9, n=28 \\
\hspace{-2.2mm} (f36  &  987.372 &  1012.79 &	9.6~ & --- & ---)& \\
\hspace{-2.2mm} (f37  &  985.603 &  1014.61 &	9.1~ & --- & ---)& \\
                 f38  &  982.680 &  1017.63 &  11.0~ &   6 &  19 & TI; MR \\
                 f39  &  979.808 &  1020.61 &  10.1~ & --- & --- & \\
                 f40  &  972.769 &  1027.99 &	9.5~ & --- & --- & \\
                 f41  &  968.027 &  1033.03 &  18.5~ &  12 &  39 & TI; NoR \\
                 f42  &  957.031 &  1044.90 &  16.5~ & --- & --- & \\
                 f43  &  956.670 &  1045.29 &  24.0~ & --- & --- & MR \\
\hspace{-2.2mm} (f44  &  949.703 &  1052.96 &	8.4~ & --- & ---)& \\
                 f45  &  945.282 &  1057.89 &  13.7~ & --- & --- & \\
\hspace{-2.2mm} (f46  &	 942.577 &  1060.92 &	8.8~ & --- & ---)& \\ % NEW !!!!!!!!!!!
                 f47  &  942.061 &  1061.50 &  36.7~ &  12 &  40 & TI \\
                 f48  &  919.003 &  1088.14 &  34.8~ &  12 &  41 & or l=7, n=24; SR \\ %; |f48-(3f130+f157)|=0.092\muHz
                 f49  &  912.554 &  1095.83 &  12.6~ & --- & --- & \\
                 f50  &  904.752 &  1105.28 &  39.3~ &  10 &  34 & TI; NoR \\
\hspace{-2.2mm} (f51  &  902.985 &  1107.44 &	9.4~ & --- & ---)& \\
\hspace{-2.2mm} (f52  &  896.904 &  1114.95 &	8.4~ &   9 &  31)& or l=12, n=42 \\
                 f53  &  896.305 &  1115.69 &  14.5~ &  12 &  42 & or l=9, n=31 \\
                 f54  &  894.152 &  1118.38 &  11.8~ & --- & --- & \\
\hspace{-2.2mm} (f55  &  890.156 &  1123.40 &	9.5~ &   8 &  27)& TI \\
                 f56  &  885.065 &  1129.86 &  20.7~ &   6 &  21 & TI MR \\
                 f57  &  881.686 &  1134.19 &  26.6~ &   7 &  25 & TI \\
                 f58  &  879.016 &  1137.64 &  11.0~ &  10 &  35 & TI \\
                 f59  &  877.037 &  1140.20 &  12.8~ &  12 &  43 & TI \\
                 f60  &  867.057 &  1153.33 &  11.7~ &   9 &  32 & TI \\ %; |f60-(2f132+2f137)|=0.083\muHz
                 f61  &  864.231 &  1157.10 &  36.8~ & --- & --- & MR \\
\hline
\end{tabular}
\label{tab1}
\end{table*}

%\newpage

\begin{table*} 
\centering
\contcaption{}
\begin{tabular}{lrrrrll}
\hline
{\bf ID} & \multicolumn{1}{c}{\bf F} & \multicolumn{1}{c}{\bf P} & \multicolumn{1}{c}{\bf A}
   & \multicolumn{1}{c}{\bf l} & \multicolumn{1}{c}{\bf n} & {\bf Notes}\\
   & \multicolumn{1}{c}{[\muHz]} & \multicolumn{1}{c}{[s]} 
   & \multicolumn{1}{c}{[ppm]} &    &	  & \\
\hline
\hspace{-2.2mm} (f62  &  857.000 &  1166.86 &	9.5~ &   8 &  28)& TI \\
                 f63  &  855.757 &  1168.56 &  11.1~ &  12 &  44 & or l=10, n=36 \\
                 f64  &  853.968 &  1171.00 &  21.2~ &  10 &  36 & or l=12, n=44 \\
                 f65  &  841.157 &  1188.84 &  18.5~ &   9 &  33 & TI \\
\hspace{-2.2mm} (f66  &	 836.247 &  1195.82 &	9.6~ & --- & ---)& \\ % NEW !!!!!!!!!!!
                 f67  &  835.861 &  1196.37 &  11.0~ & --- & --- & \\
                 f68  &  835.416 &  1197.01 &  24.4~ &  12 &  45 & TI \\
                 f69  &  828.161 &  1207.49 &  13.4~ &   8 &  29 & or l=10, n=37 \\
\hspace{-2.2mm} (f70  &  823.361 &  1214.53 &	8.7~ & --- & ---)& \\
                 f71  &  818.056 &  1222.41 &  33.8~ &  12 &  46 & or l=9, n=34; or l=7, n=27\\
\hspace{-2.2mm} (f72  &  808.031 &  1237.58 &	8.8~ &  10 &  38)& or l=6, n=23 \\
\hspace{-2.2mm} (f73  &  798.261 &  1252.72 &	8.8~ &   8 &  30)& or l=12, n=47, SR \\
                 f74  &  793.296 &  1260.56 &  14.5~ &   9 &  35 & TI \\
                 f75  &  785.720 &  1272.72 &  19.6~ &   7 &  28 & TI \\
\hspace{-2.2mm} (f76  &  771.070 &  1296.90 &	8.0~ & --- & ---)& \\ % NEW !!!!!!!!!!!
                 f77  &  770.607 &  1297.68 &  10.7~ &   9 &  36 & TI, SR \\
\hspace{-2.2mm} (f78  &  770.277 &  1298.23 &	8.4~ & --- & ---)& \\ % |f78-(2f130+2f153)|=0.149\muHz \\ % NEW !!!!!!!!!!!
                 f79  &  757.778 &  1319.65 &  19.5~ &   7 &  29 & TI \\
                 f80  &  756.461 &  1321.95 &  10.7~ & --- & --- & \\
\hspace{-2.2mm} (f81  &  752.619 &  1328.69 &	8.3~ & --- & ---)& \\
                 f82  &  741.650 &  1348.34 &  24.2~ &   6 &  25 & TI \\
                 f83  &  734.678 &  1361.14 &  19.1~ &  12 &  51 & TI \\ % ; |f83-(f109+2f153)|=0.003\muHz \\
                 f84  &  732.627 &  1364.95 &  28.8~ & --- & --- & \\ % |f84-(3f137+f146)|=0.018\muHz \\
                 f85  &  732.476 &  1365.23 &  52.4~ &   7 &  30 & TI \\
\hspace{-2.2mm} (f86  &  707.210 &  1414.01 &	8.6~ &  12 &  53)& or l=7, n=31 \\
\hspace{-2.2mm} (f87  &  702.790 &  1422.90 &	8.8~ &   8 &  34)& TI \\
\hspace{-2.2mm} (f88  &  700.061 &  1428.45 &	8.8~ & --- & ---)& \\
                 f89  &  682.352 &  1465.52 &  13.3~ &   8 &  35 & TI \\
                 f90  &  681.253 &  1467.88 &  11.6~ &  12 &  55 & TI \\
                 f91  &  679.234 &  1472.25 &  10.8~ &  10 &  45 & TI \\
                 f92  &  645.689 &  1548.73 &  19.4~ &   8 &  37 & or l=12, n=58; PLC$^1$ \\
\hspace{-2.2mm} (f93  &  634.943 &  1574.95 &	9.0~ &  12 &  59)& or l=6, n=29 \\
\hspace{-2.2mm} (f94  &  625.742 &  1598.10 &	7.3~ &   7 &  35)& TI \\
                 f95  &  622.754 &  1605.77 &  14.0~ &  10 &  49 & or l=12, n=60 \\ % ; |f95-(2f137+2f153)|=0.068\muHz \\
                 f96  &  613.476 &  1630.05 &  13.1~ &   9 &  45 & or l=12, n=61 or l=6, n=30\\
                 f97  &  611.943 &  1634.14 &  26.4~ &   8 &  39 & \\
\hspace{-2.2mm} (f98  &  604.148 &  1655.22 &  10.2~ &  12 &  62)& TI \\
                 f99  &  593.813 &  1684.03 &  16.4~ &   6 &  31 & or l=12, n=63 \\
                 f100 &  593.579 &  1684.70 &  28.1~ &  12 &  63 & or l=6, n=31; MR \\
                 f101 &  581.916 &  1718.46 &  20.6~ &   8 &  41 & \\
                 f102 &  567.798 &  1761.19 &  17.4~ &   8 &  42 & \\
                 f103 &  557.560 &  1793.53 &  58.1~ &   6 &  33 & SR \\
                 f104 &  541.799 &  1845.70 &  12.8~ &   8 &  44 & \\
                 f105 &  540.767 &  1849.23 &  31.4~ &  12 &  69 & TI; MR \\
                 f106 &  540.551 &  1849.97 &  45.7~ &   6 &  34 & \\ % |f106-(f130+3f161)|=0.121\muHz \\
                 f107 &  525.061 &  1904.54 &  72.5~ &   6 &  35 & \\
                 f108 &  524.847 &  1905.32 &  39.0~ &  10 &  58 & TI \\
                 f109 &  510.027 &  1960.68 & 164.4~ &   6 &  36 & SR \\
\hspace{-2.2mm} (f110 &  503.468 &  1986.22 &  11.5~ &   5 &  31)& \\ % AMPL SEEMS OVEREST. !
\hspace{-2.2mm} (f111 &  495.961 &  2016.29 &	8.1~ &   6 &  37)& \\ % |f111-(3f146+f161)|=0.033\muHz, |f111-(2f146+2f153)|=0.137\muHz \\ % NEW !
                 f112 &  487.276 &  2052.22 &  36.0~ &   5 &  32 & MR \\ % ; |f112-(2f137+2f161)|=0.008\muHz \\
\hspace{-2.2mm} (f113 &  482.714 &  2071.62 &	9.5~ &   6 &  38)& \\
                 f114 &  472.533 &  2116.25 &  35.5~ &   5 &  33 & MR \\ % ; |f114-(f146+3f153)|=0.027\muHz \\
                 f115 &  470.009 &  2127.62 &  42.1~ &   6 &  39 & MR \\ % ; |f115-(f132+3f165)|=0.066\muHz \\
                 f116 &  458.005 &  2183.38 &  33.6~ &   5 &  34 & or l=6, n=40; or l=4, n=28\\
                 f117 &  446.757 &  2238.35 &  69.3~ &   6 &  41 & MR \\ % potential LC 
                 f118 &  445.031 &  2247.03 &  23.0~ &   5 &  35 & MR \\
                 f119 &  441.287 &  2266.10 &  38.6~ &   4 &  29 & NoR \\
                 f120 &  432.204 &  2313.72 &  26.2~ &   5 &  36 & MR \\
                 f121 &  409.048 &  2444.70 &  52.0~ &   5 &  38 & MR \\ % potential LC
                 f122 &  399.031 &  2506.07 &  41.8~ &   4 &  32 & MR \\
                 f123 &  386.847 &  2585.00 &  41.3~ &   4 &  33 & MR \\ % potential LC
\hline
\end{tabular}
\end{table*}

\begin{table*} 
\centering
\contcaption{}
\begin{tabular}{lrrrrll}
\hline
{\bf ID} & \multicolumn{1}{c}{\bf F} & \multicolumn{1}{c}{\bf P} & \multicolumn{1}{c}{\bf A}
   & \multicolumn{1}{c}{\bf l} & \multicolumn{1}{c}{\bf n} & {\bf Notes}\\
   & \multicolumn{1}{c}{[\muHz]} & \multicolumn{1}{c}{[s]} 
   & \multicolumn{1}{c}{[ppm]} &    &	  & \\
\hline
                 f124 &  384.362 &  2601.71 &  77.3~ & --- & --- & SR \\ % [l=2, n=18] potential LC
                 f125 &  374.989 &  2666.74 &  20.3~ &   4 &  34 & NoR \\
                 f126 &  353.644 &  2827.71 &  61.6~ &   4 &  36 & MR; PLC$^2$ \\
                 f127 &  343.924 &  2907.62 &  18.9~ &   4 &  37 & \\ % potential LC
                 f128 &  325.591 &  3071.34 &  52.6~ &   4 &  39 & or l=6, n=56; NR \\ % potential LC
\hspace{-2.2mm} (f129 &  309.680 &  3229.14 &  12.6~ &   4 &  41)& or l=5, n=50 \\ % potential LC
                 f130 &  272.740 &  3666.50 & 185.5~ &   2 &  25 & NoR \\
                 f131 &  262.856 &  3804.37 &  47.0~ &   2 &  26 & \\
                 f132 &  234.551 &  4263.47 & 217.1~ &   2 &  29 & NoR \\
                 f133 &  226.883 &  4407.56 &  83.4~ & --- & --- & PLC$^3$ \\
                 f134 &  226.772 &  4409.72 &  79.5~ &   2 &  30 & \\
\hspace{-2.2mm} (f135 &  212.630 &  4703.01 &  21.6~ &   2 &  32)& \\
\hspace{-2.2mm} (f136 &  200.065 &  4998.36 &  26.3~ &   2 &  34)& or l=1, n=20 \\
                 f137 &  199.019 &  5024.65 & 103.8~ &   1 &  20 & or l=2, n=34; NR \\
                 f138 &  184.250 &  5427.40 &  52.0~ & --- & --- & PLC$^4$ \\
                 f139 &  183.842 &  5439.44 &  69.5~ &   2 &  37 & MR \\
\hspace{-2.2mm} (f140 &  178.849 &  5591.29 &  30.8~ &   2 &  38)& PLC$^5$ \\
\hspace{-2.2mm} (f141 &  174.249 &  5738.90 &  28.8~ &   2 &  39)& PLC$^6$ \\
                 f142 &  165.701 &  6034.97 &  74.8~ &   2 &  41 & \\
\hspace{-2.2mm} (f143 &  145.725 &  6862.25 &  42.8~ &   1 &  27)& \\
                 f144 &  144.346 &  6927.78 &  49.1~ &   2 &  47 & \\
                 f145 &  140.381 &  7123.45 &  68.7~ &   1 &  28 & \\
                 f146 &  135.588 &  7375.29 & 180.0~ &   2 &  50 & or l=1, n=29 \\
                 f147 &  126.811 &  7885.73 &  55.7~ &   1 &  31 & \\
                 f148 &  125.496 &  7968.39 & 163.3~ &   2 &  54 & \\
                 f149 &  123.141 &  8120.80 &  75.9~ &   2 &  55 & or l=1, n=32 \\
                 f150 &  119.146 &  8393.06 & 355.3~ &   1 &  33 & or l=2, n=57; MR \\
                 f151 &  115.962 &  8623.53 & 131.4~ &   1 &  34 & \\
                 f152 &  115.645 &  8647.18 &  62.5~ & --- & --- & \\
                 f153 &  112.324 &  8902.79 & 223.3~ &   1 &  35 & MR \\
                 f154 &  106.398 &  9398.65 & 150.4~ &   1 &  37 & SR \\ 
                 f155 &  102.991 &  9709.61 &  73.2~ &   1 &  38 & TI \\
                 f156 &  102.856 &  9722.33 &  70.1~ &   2 &  66 & TI \\
                 f157 &  100.875 &  9913.29 & 502.5~ &   1 &  39 & SR \\
                 f158 &   95.776 & 10440.98 & 457.4~ &   1 &  41 & MR \\
\hspace{-2.2mm} (f159 &   92.884 & 10766.12 &  50.9~ &   2 &  73)& \\
                 f160 &   91.298 & 10953.08 & 322.4~ &   1 &  43 & SR \\
                 f161 &   89.230 & 11207.02 & 146.6~ &   1 &  44 & \\
                 f162 &   85.032 & 11760.31 &  81.3~ &   1 &  46 & TI \\
                 f163 &   83.396 & 11990.97 & 796.1~ &   1 &  47 & NoR \\
                 f164 &   80.035 & 12494.57 & 231.9~ &   1 &  49 & \\
                 f165 &   78.508 & 12737.50 & 321.9~ &   1 &  50 & or l=2, n=86; MR \\
                 f166 &   77.842 & 12846.60 & 144.5~ &   2 &  87 & \\
                 f167 &   73.613 & 13584.56 &  74.9~ &   2 &  92 & or l=1, n=53 \\
                 f168 &   68.027 & 14700.08 &  71.5~ &   2 &  99 & or l=1, n=58 \\
                 f169 &   65.073 & 15367.46 & 306.0~ &   2 & 104 & or l=1, n=60 \\
%                 f170 &   37.302 & 26808.09 &  69.3~ & --- & --- & \\ % likely artifact 
%                                                                      % due to K2 repointings
\hline
\multicolumn{7}{l}{The overtone {\bf n} in column 6 is arbitrarily defined assuming that $n$=1}\\
\multicolumn{7}{l}{corresponds to the first positive pulsation period starting from zero}\\
\multicolumn{7}{l}{(and assuming a constant period spacing down to $n$=1).}
%\multicolumn{7}{l}{down to $n$=1).}
%\multicolumn{7}{l}{corresponds to the first positive pulsation period starting from zero.}
%\multicolumn{7}{l}{with $n$=1 is the 1st one starting from a pulsation period equal to zero.}
% assuming that the mode with $n$=1 is the 1st one starting from a pulsation period equal to zero.
% assuming that the constant-period-spacing sequence propagates down to n=1 and that $n$=1 corresponds to the first positive pulsation period starting from zero.
\vspace{1mm}
\\
\multicolumn{7}{l}{Notes: TI=Tentative Identification.}\\
\multicolumn{7}{l}{\hspace{8.4mm} SR=Strong Residuals after prewhitening 
due to amplitude/}\\ 
\multicolumn{7}{l}{\hspace{14.4mm} frequency variations and/or unresolved  
close frequencies.}\\
\multicolumn{7}{l}{\hspace{8.4mm} MR=Moderate Residuals after prewhitening 
due to potential}\\ 
\multicolumn{7}{l}{\hspace{15.5mm} ampl./freq. variations and/or 
unresolved close frequencies.}\\
\multicolumn{7}{l}{\hspace{8.4mm} NoR=No Residuals after prewhitening: single
stable peak.}\\
\multicolumn{7}{l}{\hspace{8.4mm} PLC$^1$=Potential Linear Combination: |f92-(f109+f146)|=0.074\muHz.}\\
\multicolumn{7}{l}{\hspace{8.4mm} PLC$^2$: |f126-(f132+f150)|=0.053\muHz.}\\
\multicolumn{7}{l}{\hspace{8.4mm} PLC$^3$: |f133-(f146+f160)|=0.003\muHz.}\\
\multicolumn{7}{l}{\hspace{8.4mm} PLC$^4$: |f138-(f150+f169)|=0.031\muHz, |f138-(f154+f166)|=0.010\muHz,}\\ 
\multicolumn{7}{l}{\hspace{17.7mm} |f138-(f157+f163)|=0.021\muHz.}\\
\multicolumn{7}{l}{\hspace{8.4mm} PLC$^5$: |f140-(f157+f166)|=0.132\muHz.}\\
\multicolumn{7}{l}{\hspace{8.4mm} PLC$^6$: |f141-(f158+f165)|=0.035\muHz.}\\
\end{tabular}
\end{table*}

\subsection{Pulsation frequencies}

HD\,4539 was observed for 78.7 days by $K2$ in short cadence from
BJD$_{TBD}$ 2457392.058531 to 2457470.781242 (corresponding to 04/01/2016 --
23/03/2016).
We downloaded all available data from the ``Barbara A. Mikulski Archive for
Space Telescopes'' (MAST)\footnote{archive.stsci.edu}. 
We used the short-cadence (SC) data, sampled at 58.85\,s time resolution, 
since they allow us to reasonably sample an amplitude spectrum beyond 
the g-mode region, which means that both g-and p-mode regions are covered.

First, we used standard \texttt{IRAF} tasks to extract fluxes from the pixel 
tables.
Next, we used our custom \texttt{Python} scripts to decorrelate fluxes 
in X and Y directions. This latter step removed the flux dependence on 
position on the CCD and the resultant light curve was free of the signatures 
of thruster firings. 
Finally, the light variations were converted to residual flux
%($a=f/\bar{f}-1$) 
($f/\bar{f}-1$) in 
%{\it parts per thousand} (ppt) or 
{\it parts per million} (ppm).

In order to extract the pulsation frequencies from the light curve we used 
the following procedure: first we defined the mean noise level of the amplitude
spectrum.
This was done by selecting 123 peaks higher than a certain treshold in the 
amplitude spectrum of the data, regardless of whether they were true pulsation 
frequencies or not, subtracting these frequencies from the light curve 
(pre-whitening), computing the amplitude spectrum of the residuals, and
applying to it a cubic spline interpolation.
%We took this cubic spline interpolation as the mean noise level ($\sigma$) 
%as a function of frequency.
This cubic spline interpolation represents the mean noise level ($\sigma$) 
as a function of frequency.\footnote{Note that even if 123, the number of 
selected peaks, is arbitrary, changing this number has very little influence on
the mean noise level since the spline interpolation is performed 
after dividing the amplitude spectrum of the residuals in many subsets, 
computing the mean over each subset, and requiring the spline to pass 
through these average values.}
At this point we started the real extraction of the pulsation 
frequencies assuming a 5.4\,$\sigma$ threshold, which corresponds to 
a 95\% confidence level for $K2$ data following \citet{baran15}.
A low-frequency peak at 37.3~$\mu$Hz was excluded a-priori because
the corresponding period of $\sim$26,800~s is too close to 6-7 hours,
which is the typical time between two re-pointings of the $K2$ telescope
(see e.g. Fig.~1 of \citealt{vandenburg14}).\\
The frequency extraction was done in two steps:\\
{\bf 1)} we selected from the amplitude spectrum of the data 83 
high-amplitude peaks ($>$7.0\,$\sigma$).
%($>$6.8\,$\sigma$).
Secondary peaks very close to the main peaks were excluded in this phase.
The frequencies, amplitudes and phases of these 83 main signals were
optimized using a least-square fit with 83$\times$3=249 free parameters.
This is about the maximum number of parameters for which we obtain a robust
convergence of the least-square solution.
From here on the least-square optimization of further frequencies was 
done one by one, optimizing frequency, amplitude and phase of each new 
entry but keeping fixed the frequencies of the 83 main peaks 
(but not their amplitudes and phases).
30 further frequencies were added, with lower amplitudes but still higher 
than 5.4\,$\sigma$.
Furthermore, we added also 11 ``close frequencies'' bringing the total 
number of robust detections to 124 (see Table~\ref{tab1}).
As ``close frequencies'' we mean those frequencies that are close 
to one of the previous 113 (83+30) peaks, but distant enough to be 
resolved at half of their maximum amplitude (half maximum). 
Frequencies not resolved at half maximum are not considered 
in Table~\ref{tab1}.
Again, frequencies, amplitudes and phases of the close frequencies were
optimized keeping fixed the frequencies of the 113 previous peaks
(but not their amplitudes and phases).\\
{\bf 2)} From the amplitude spectrum of the residuals (light curve -- 124 
frequencies) we selected 45 further peaks with an amplitude higher 
than 4.0\,$\sigma$, which corresponds to about 50\% confidence level 
\citep{baran15}, and we verified that their amplitude was higher than 
4.0\,$\sigma$ also in the amplitude spectrum of the data.
These 45 low-amplitude frequencies are considered only as candidates and
are shown in brackets in Table~\ref{tab1}.
An example of our 2-step procedure to extract the pulsation frequencies
is shown in Fig.~\ref{fig1}.

\subsection{Frequency multiplets, inclination, stellar rotation}

In the amplitude spectrum of HD\,4539 we do not see any signature of 
multiplets of equally spaced frequencies, suggesting a very long rotation 
period, longer than the $K2$ light curve, or a very low inclination of the 
rotation axis with respect to the line of sight or both.
The rotation velocity of HD\,4539 was measured by \citet{geier12}, who
obtained v$_{\rm rot}$\,sin$i$=3.9$\pm$1.0~km/s, using 21 absorption lines
from FEROS spectra.
If we assume a stellar radius of 0.26$\pm$0.01~\rsun\ (see section 3), 
this measurement translates into 
P$_{\rm rot}$/sin$i$=3.4$\pm^{1.3}_{0.8}$~days, corresponding
to a rotational frequency splitting of $\sim$1.7/sin$i$~\muHz\ ($l$=1) or 
2.8/sin$i$~\muHz\ ($l$=2).
With a high inclination these frequency splittings would be easily seen
given the $\sim$0.15~\muHz\ frequency resolution of the $K2$ data.
With a low inclination the frequency splittings would be even larger but the 
amplitude of the $m$$\neq$0 modes would be lower.
However, given the high quality of the $K2$ data, some of the low-amplitude 
components of the multiplets should be visible, e.g. the $m$=$\pm$1 
components of the $l$=2 modes \citep{pesnell85}.
Therefore, if we want to reconcile the absence of frequency splittings with
the spectroscopic result of \citet{geier12}, the only possibility would be 
an extremely low inclination and a very short rotation period of the order of 
hours.
But very short rotation periods are unusual in sdB pulsators and seem 
to happen only in close binaries \citep{reed18, charpinet18}.
Given that HD\,4539 does not have close companions (see section 4), we 
conclude that the rotation velocity of 3.9/sin$i$~km/s should be considered 
just as an upper limit.
This interpretation is also supported by the fact that pulsational line 
broadening is certainly present in this star.
The RV variations of several hundred meters per second that we measured 
(section 4) include both vertical and horizontal motions. 
But we know that the horizontal velocity field is dominant in g-modes so that 
the horizontal velocities, whose contribution is maximum at the limb of the 
star, just as the rotation does, may easily reach several km/s.
In conclusion: 1) we do not have reasons to believe that the rotation period 
is much different from the relatively long periods found in almost all the 
other sdB pulsators.
2) This star must have a very long rotation period and/or a very low 
inclination. This is further confirmed by the fact that some of the main peaks
in the amplitude spectrum (those indicated with ``NoR'' in the last column of 
Table~\ref{tab1}) are completely removed when subtracting a single sine wave, 
indicating that there are no unresolved multiplets in these cases.

\subsection{Period spacing}

\begin{figure}
\centering
\includegraphics[width=8.6cm,angle=0]{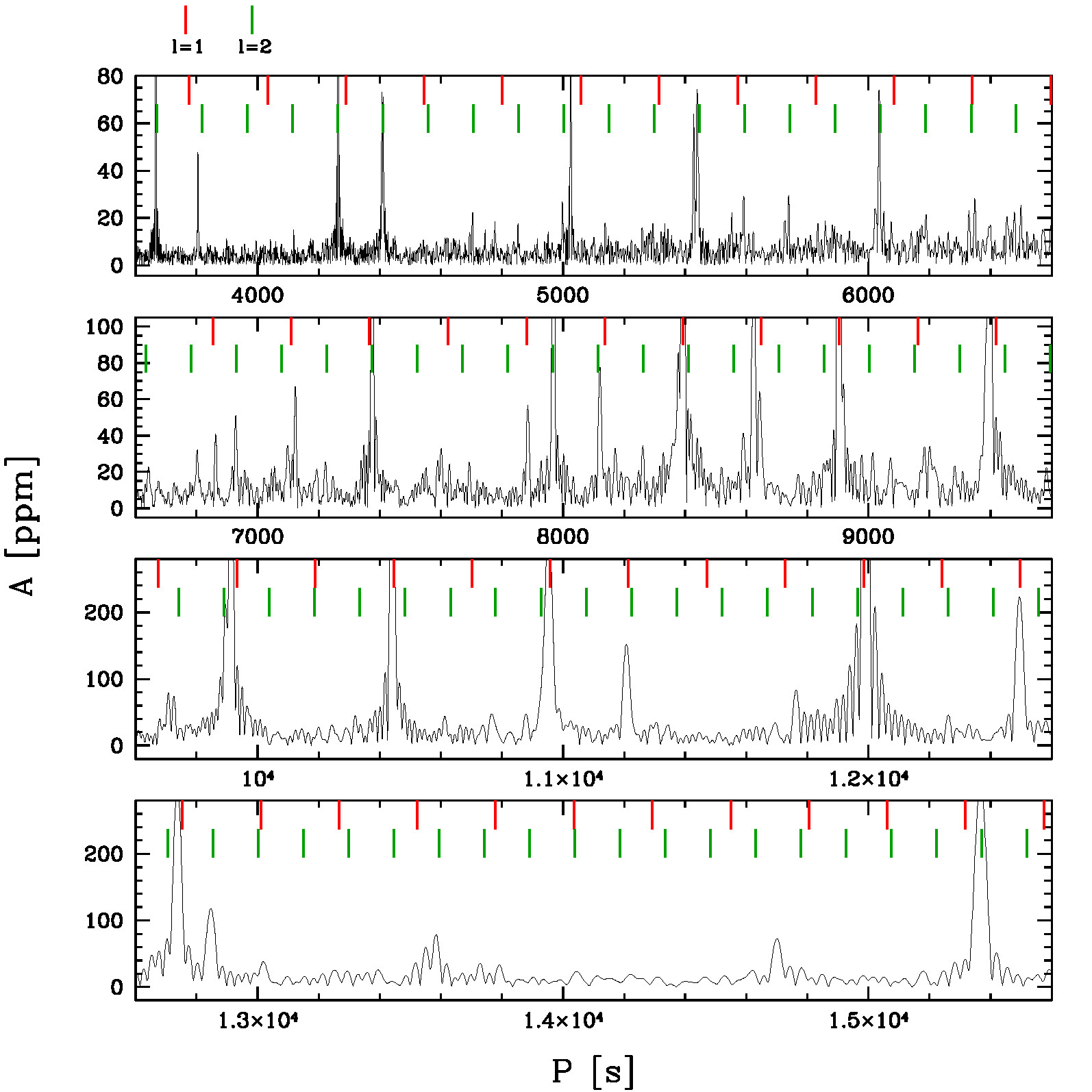}
\vspace{-3mm}
\caption{Period spacing for the modes with $l$=1,2: 
both sequences are well defined.}
\label{fig2}
\end{figure}

\begin{figure}
\centering
\includegraphics[width=8.6cm,angle=0]{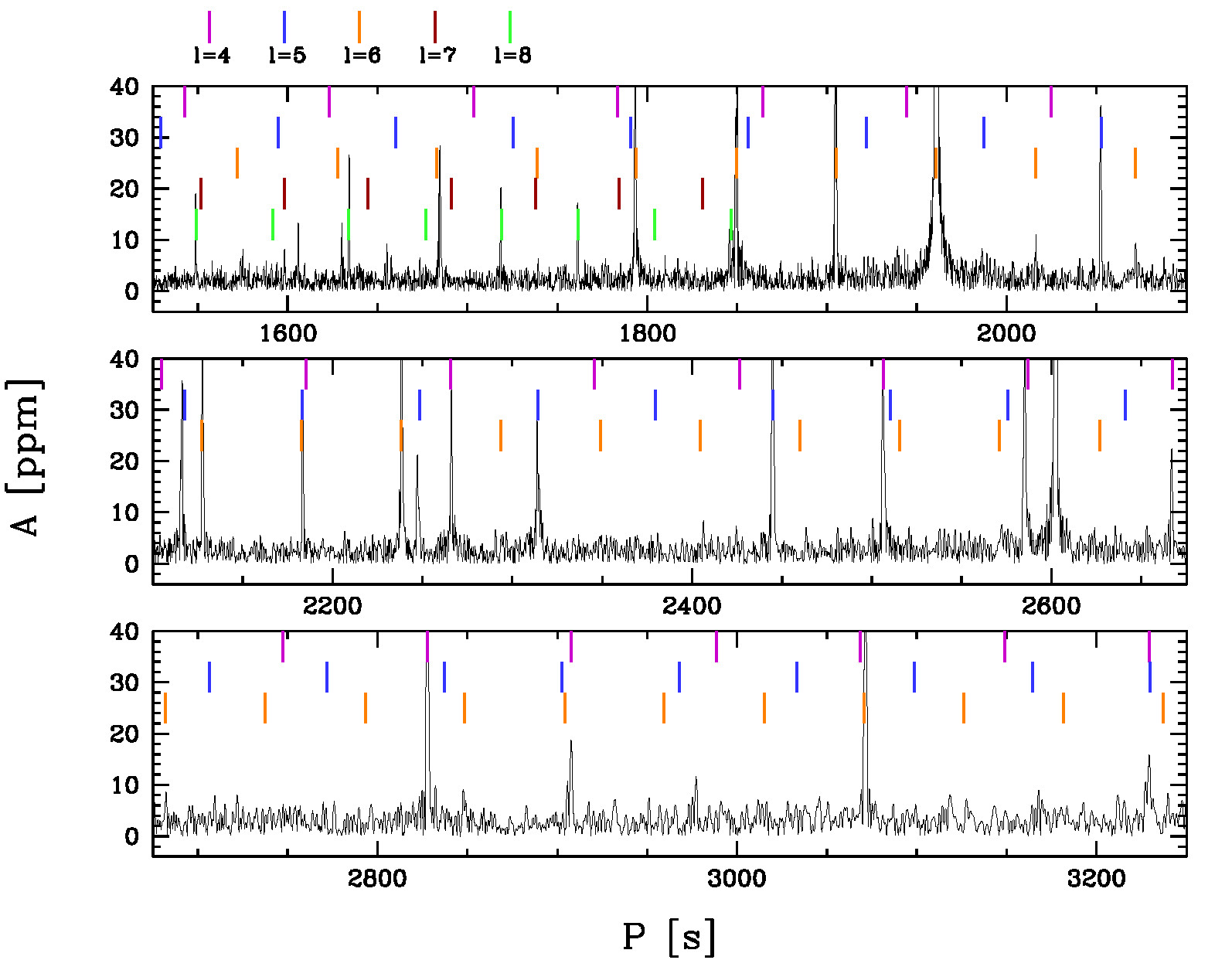}
\vspace{-3mm}
\caption{Period spacing for the modes with 4$\le$$l$$\le$8. Note the clean
sequence of modes with $l$=4 (between $\sim$2260 and $\sim$3230\,s), $l$=5 
($\sim$2050--2450\,s, with 5 consecutive modes), $l$=6 ($\sim$1680--2240\,s,
with at least 9 consecutive modes) and $l$=8 ($\sim$1550--1850\,s). 
The modes with $l$=7 are not active in this region and can be seen at shorter 
periods in Fig.~\ref{fig4}.}
\label{fig3}
\end{figure}

\begin{figure}
\centering
\includegraphics[width=8.6cm,angle=0]{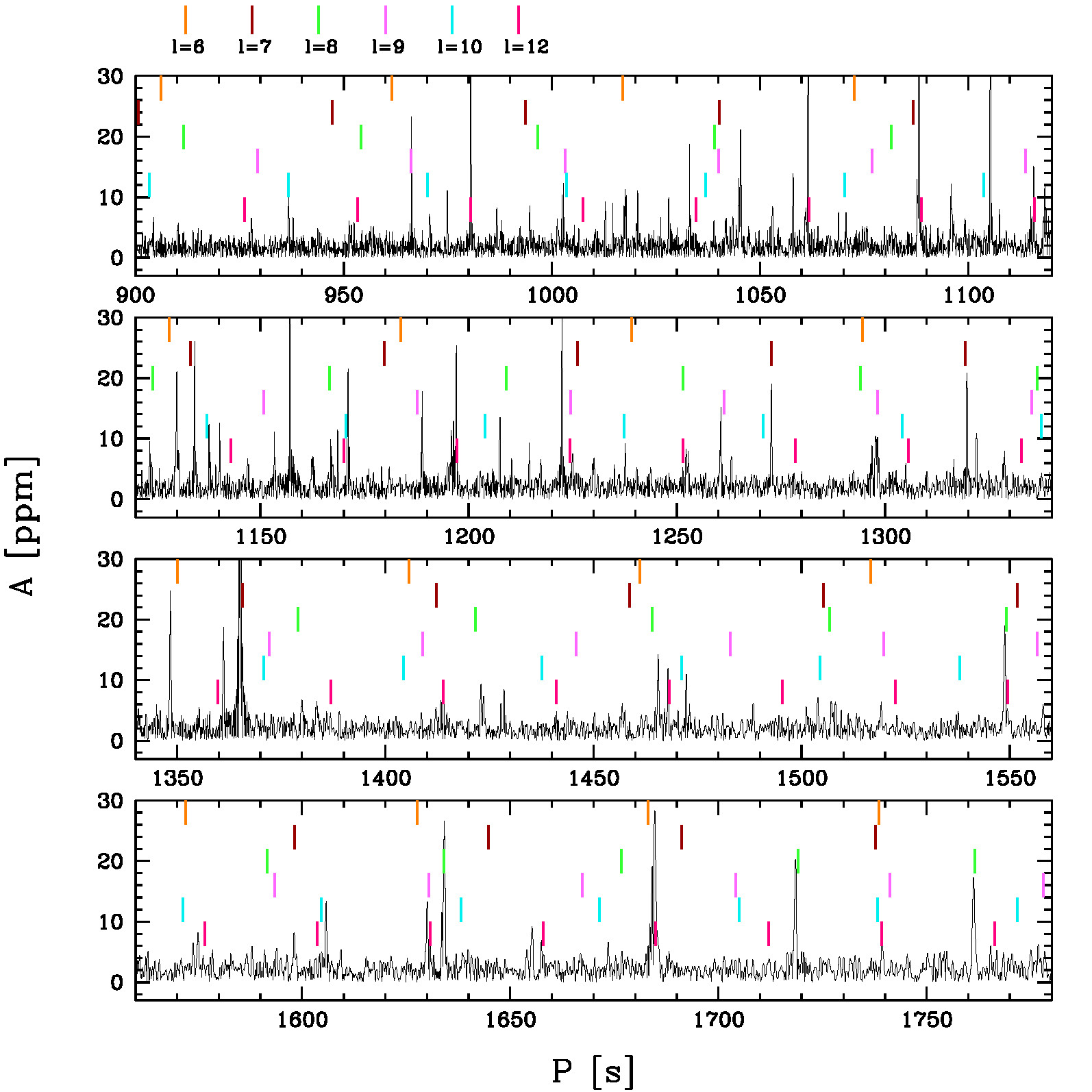}
\vspace{-3mm}
\caption{Period spacing for the modes with 6$\le$$l$$\le$12.
The high concentration of peaks at short periods implies that modes with high 
degree, up to at least $l$=12, must be present in this star.
However it makes the mode identification very difficult. 
Sequences with at least 2 or 3 consecutive peaks can be recognized for
$l$=7 ($\sim$1090--1370\,s) and $l$=8($\sim$1550--1760\,s, see also 
Fig.~\ref{fig3}), and partially also for $l$=9 ($\sim$1190--1300\,s), $l$=10 
($\sim$905--1240\,s), and $l$=12 ($\sim$980 to at least 1200).}
\label{fig4}
\end{figure}

\begin{figure*}
\centering
%%%%%%%%
\begin{minipage}[c]{8.8cm}
\hspace{22.5mm}
\includegraphics[height=5.6cm]{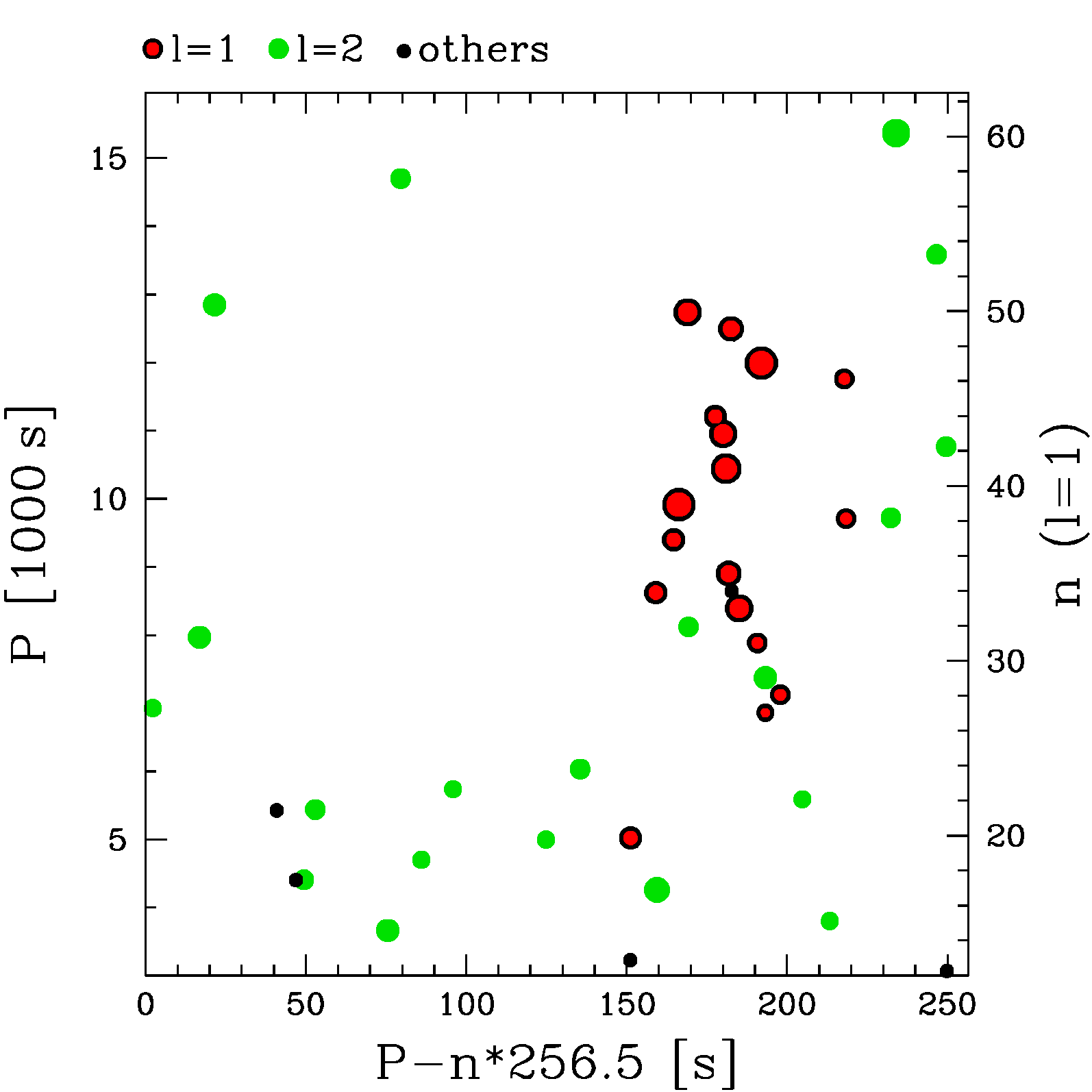}
\end{minipage}
\begin{minipage}[c]{8.8cm}
\hspace{8mm}
\includegraphics[height=5.6cm]{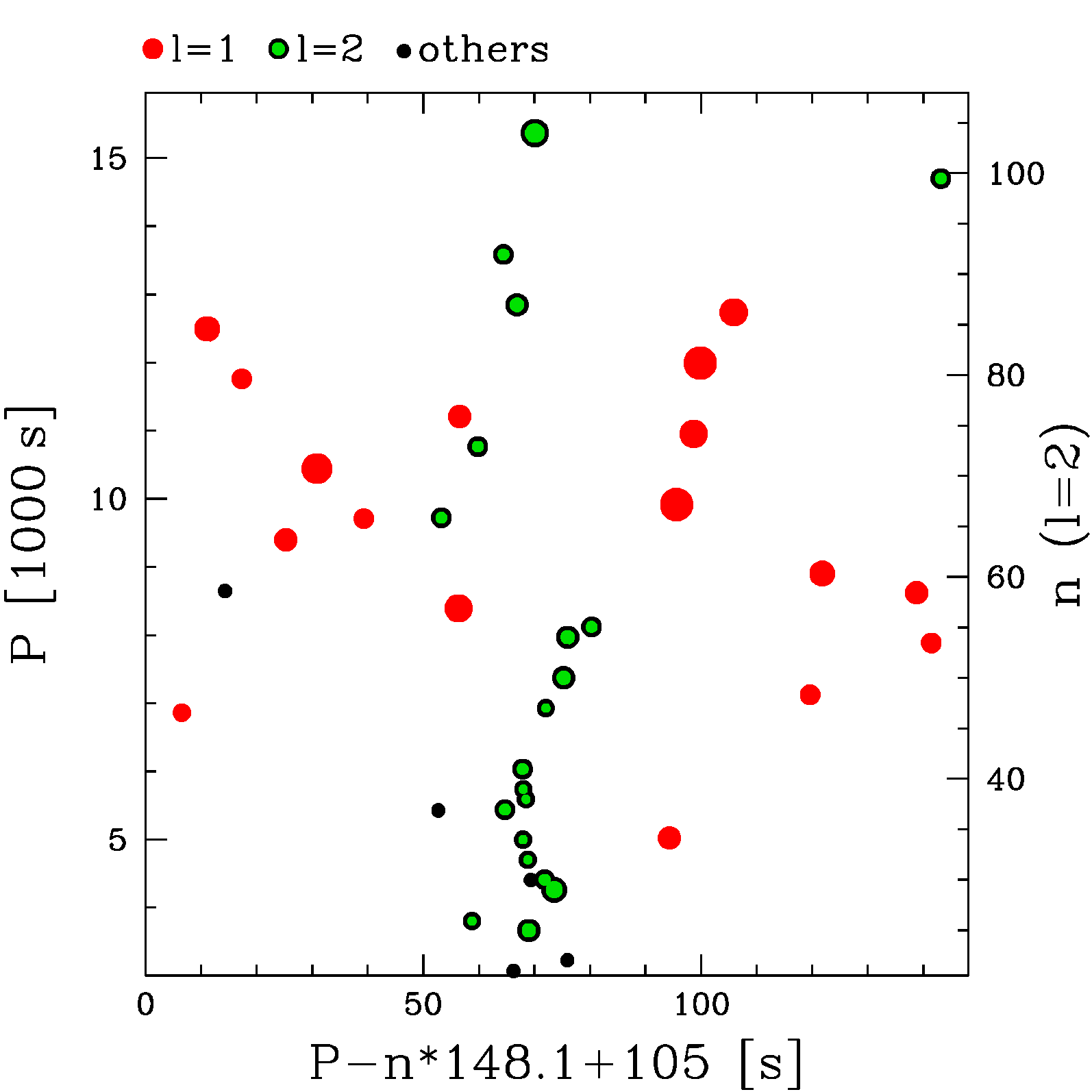}
\end{minipage}
%%%%%%%%
\begin{minipage}[l]{5.84cm}
\includegraphics[height=5.6cm]{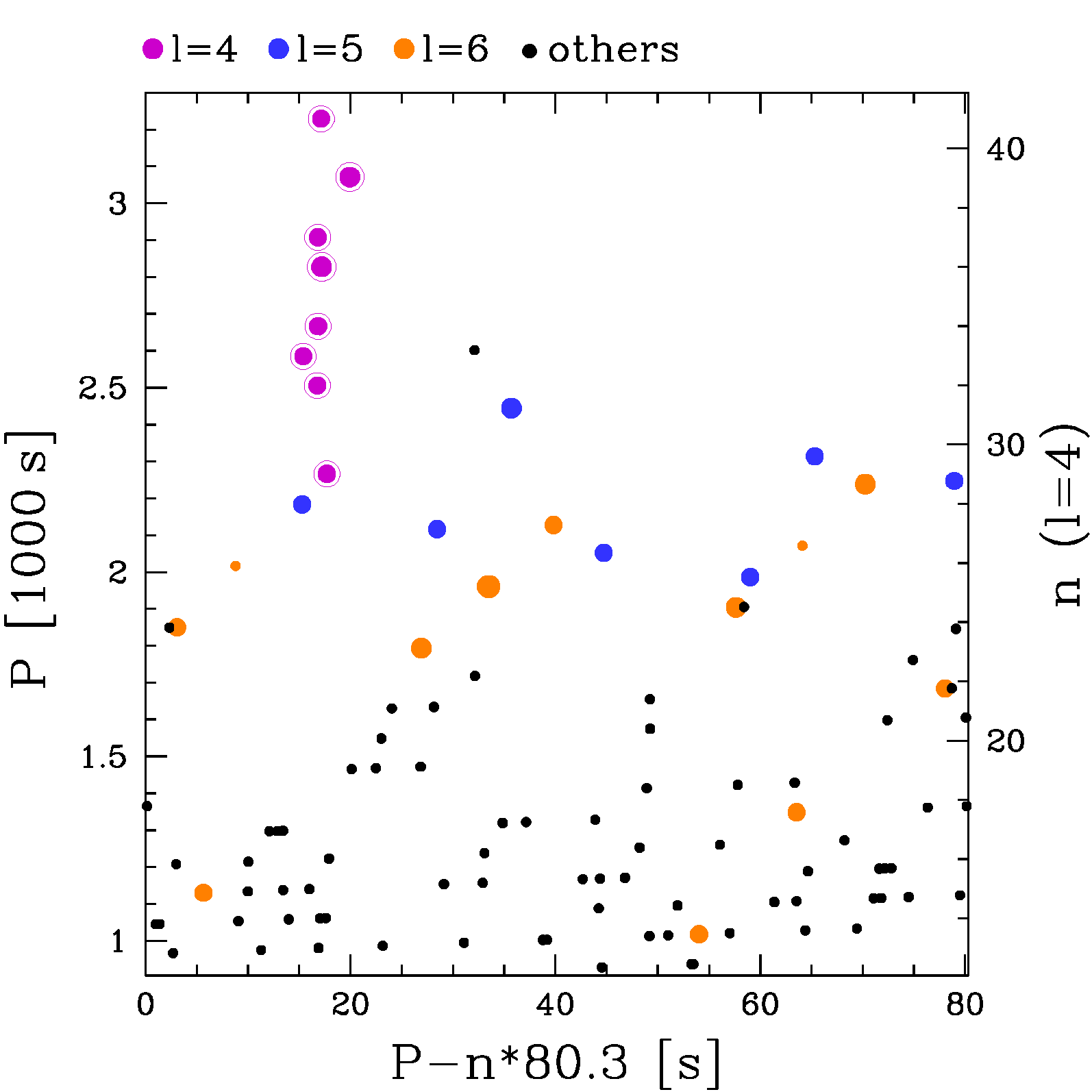}
\end{minipage}
\begin{minipage}[c]{5.84cm}
\includegraphics[height=5.6cm]{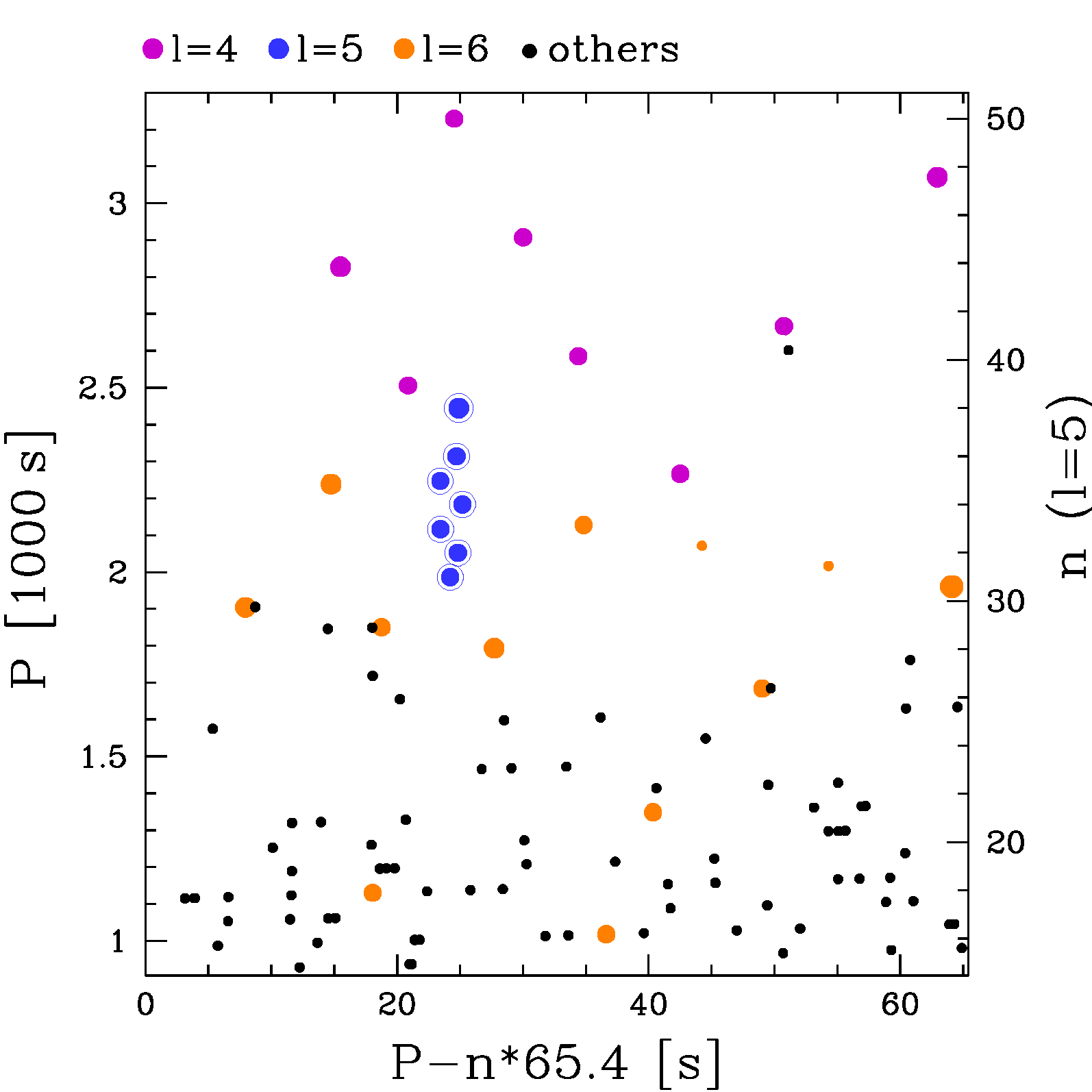}
\end{minipage}
\begin{minipage}[r]{5.84cm}
\includegraphics[height=5.6cm]{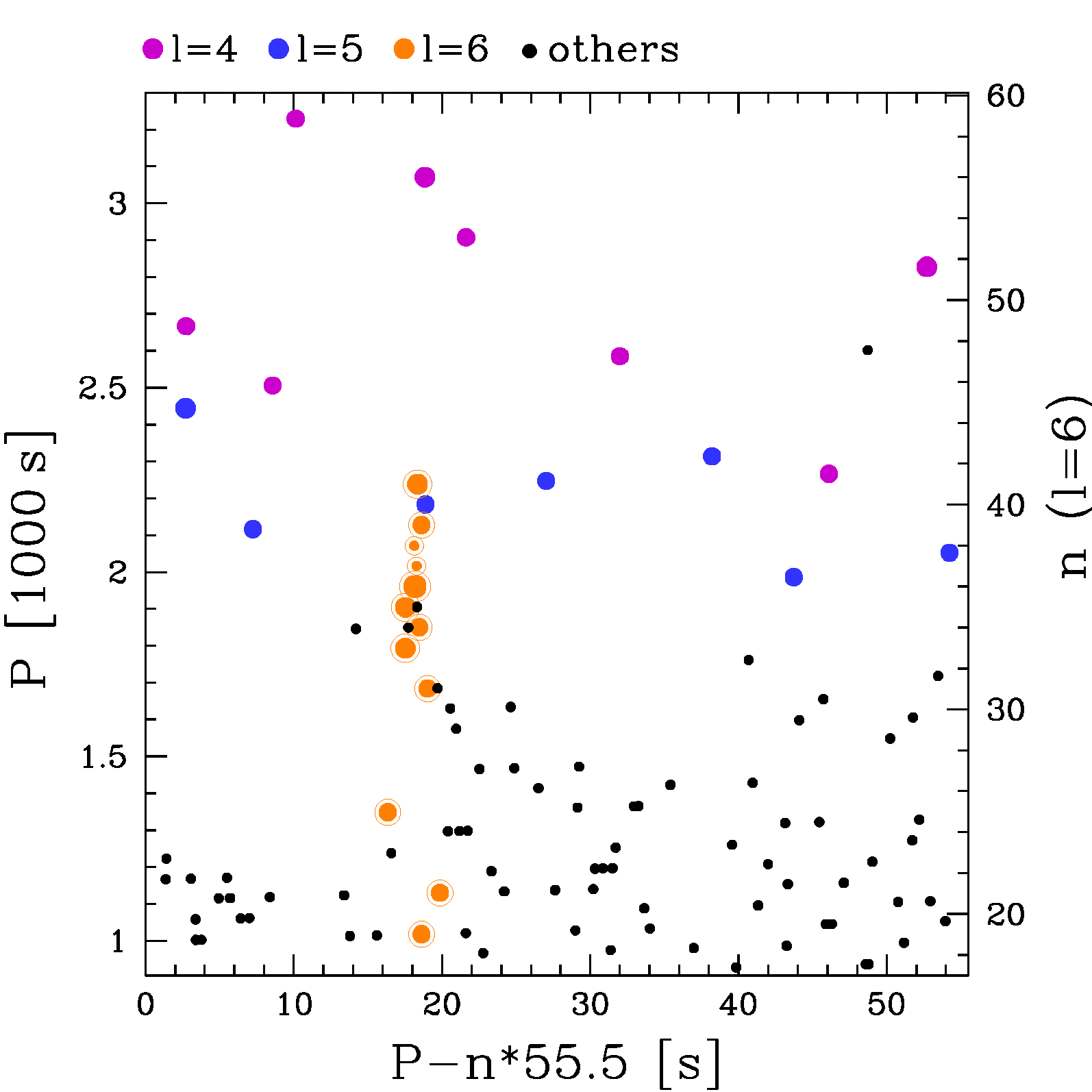}
\end{minipage}
%%%%%%%%
\begin{minipage}[r]{5.84cm}
\includegraphics[height=5.6cm]{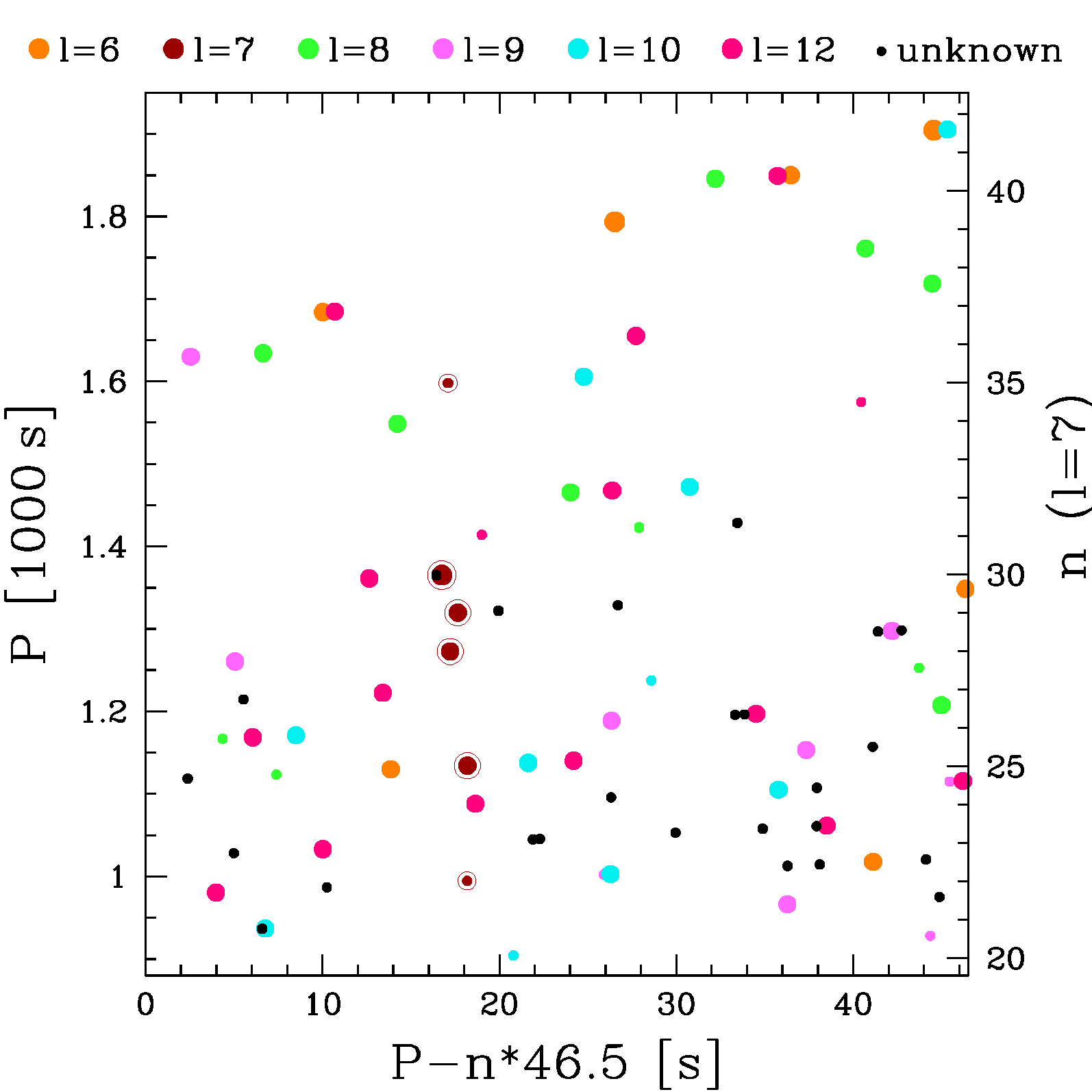}
\end{minipage}
\begin{minipage}[r]{5.84cm}
\includegraphics[height=5.6cm]{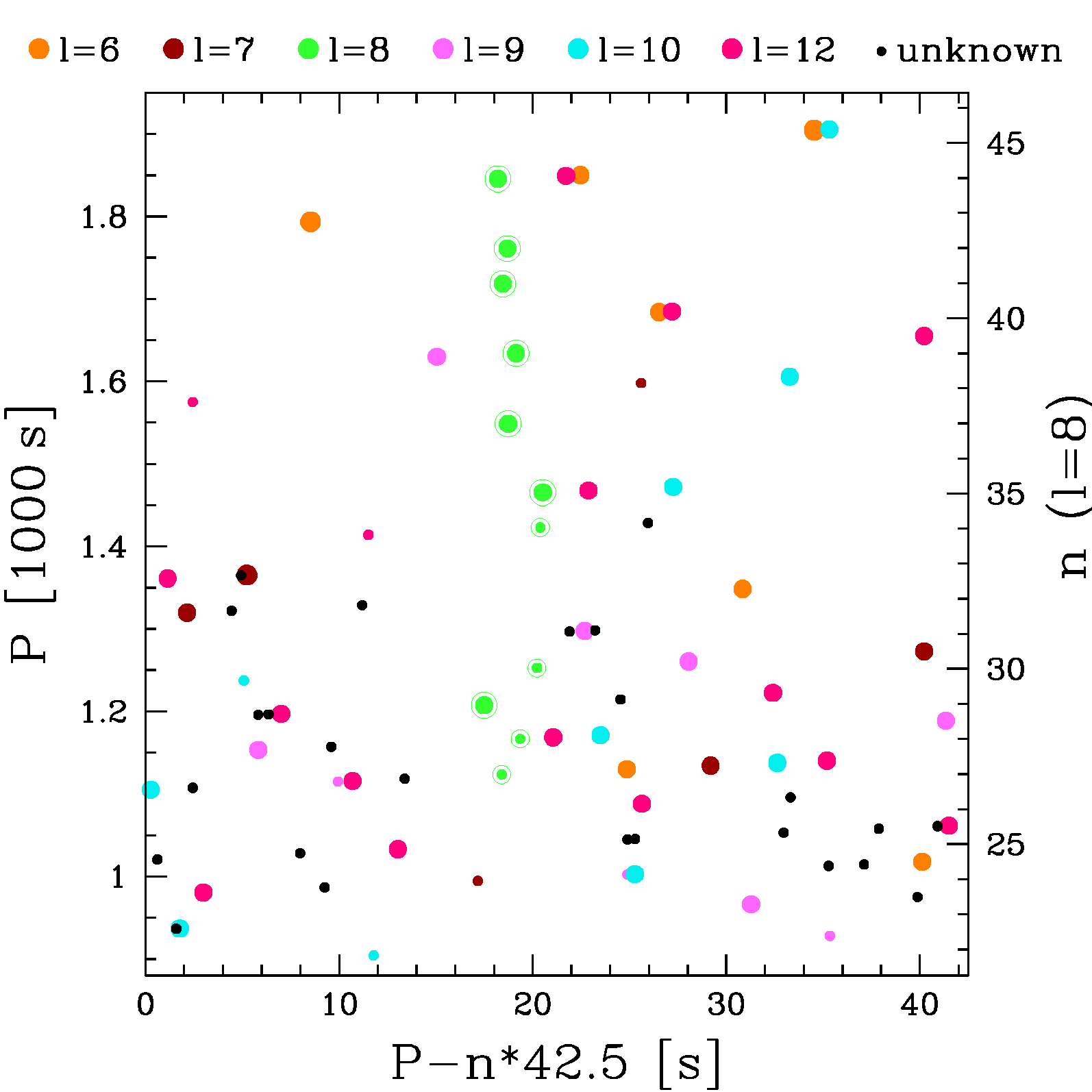}
\end{minipage}
\begin{minipage}[r]{5.84cm}
\includegraphics[height=5.6cm]{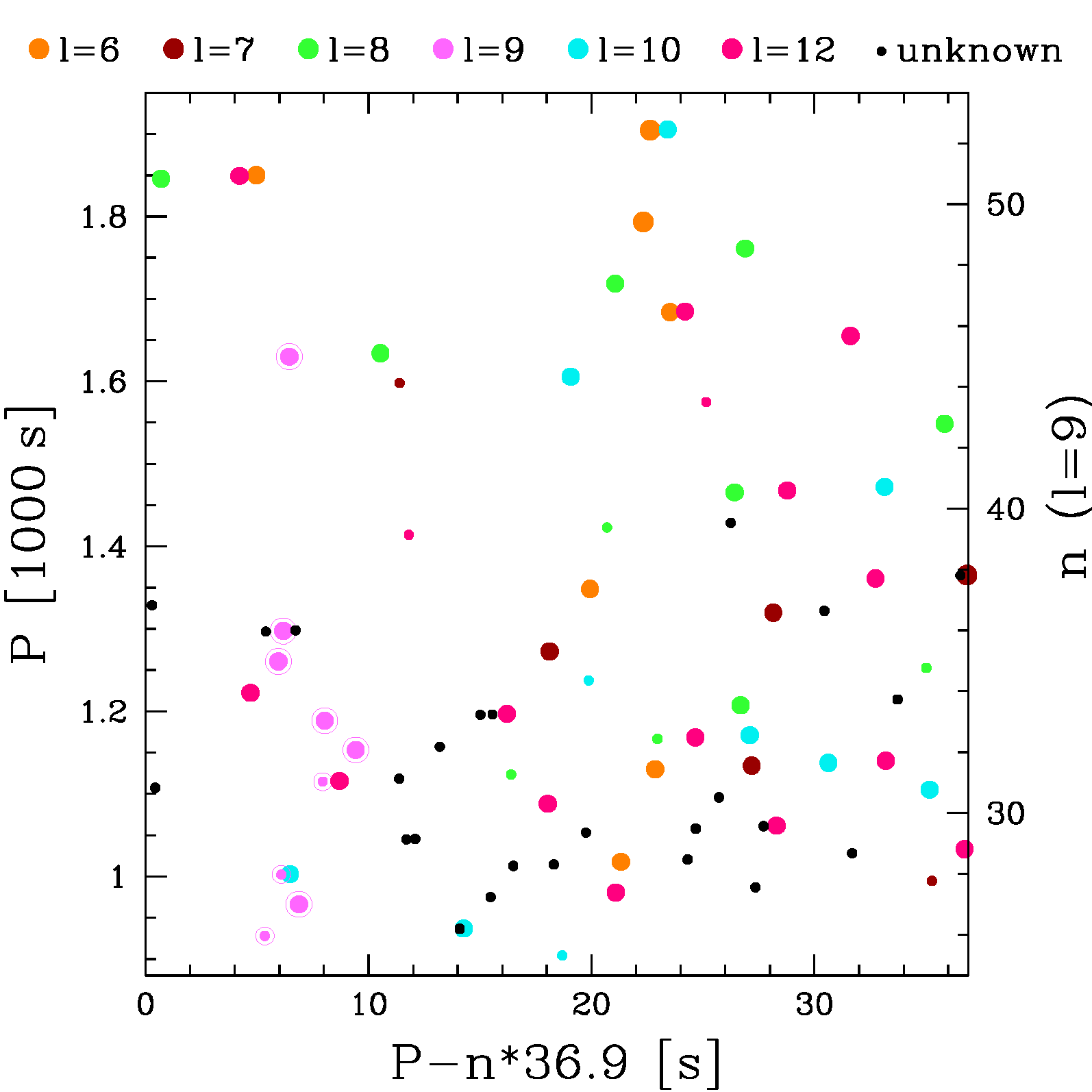}
\end{minipage}
%%%%%%%%
\begin{minipage}[c]{8.8cm}
\hspace{22.5mm}
\includegraphics[height=5.6cm]{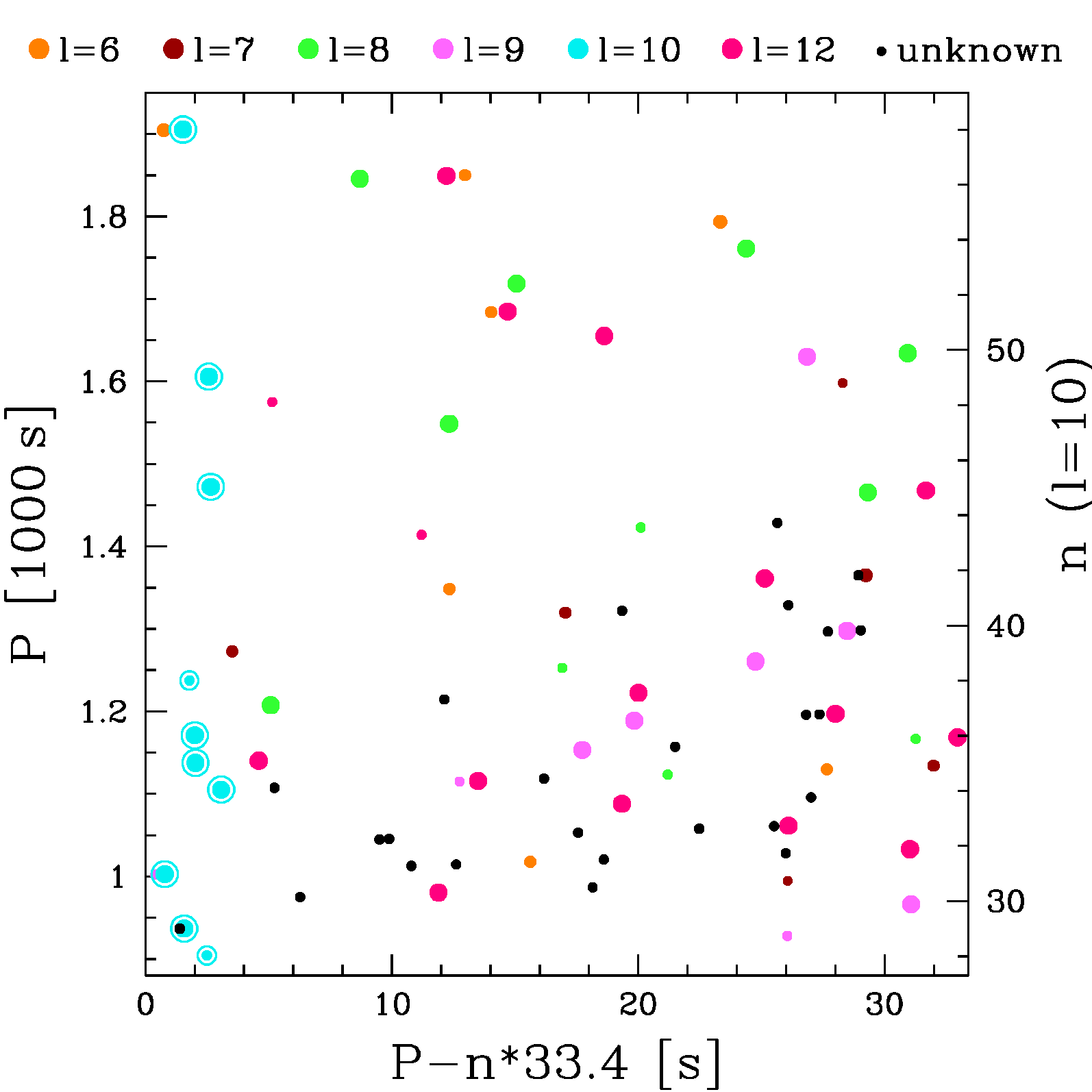}
\end{minipage}
\begin{minipage}[c]{8.8cm}
\hspace{8mm}
\includegraphics[height=5.6cm]{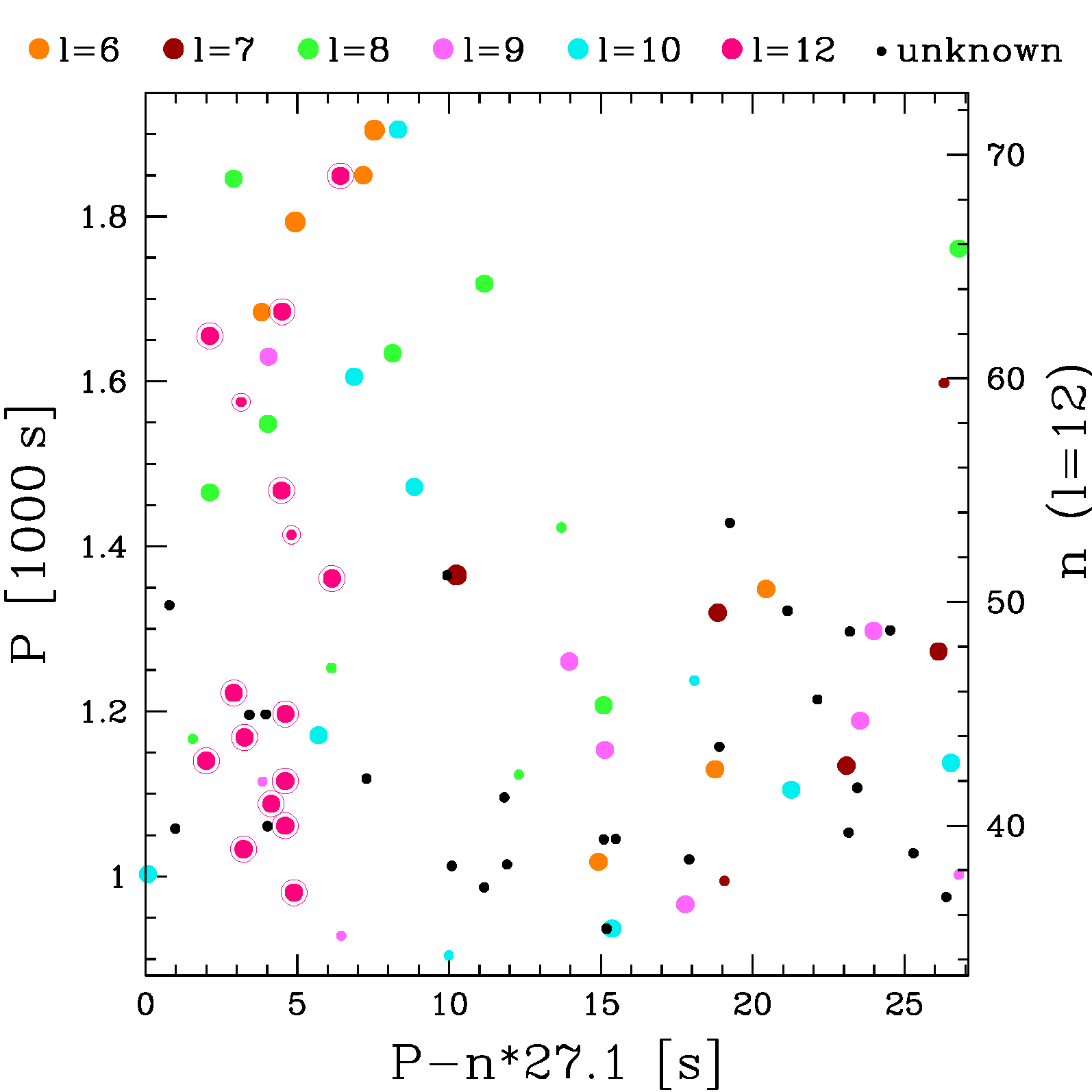}
\end{minipage}
%%%%%%%%
\vspace{1mm}
\caption{Echelle diagrams of the modes with $l$=1,2,4,5,6,7,8,9,10,12.
The size of each point is proportional to the amplitude of that mode.
An offset of 105\,s was applied to the upper right panel ($l$=2)
just for clarity.
%For the modes with $l$$>$2 the points have been highlighted for clarity.
}
\label{fig5}
\end{figure*}

\begin{figure}
\centering
\includegraphics[width=8.45cm,angle=0]{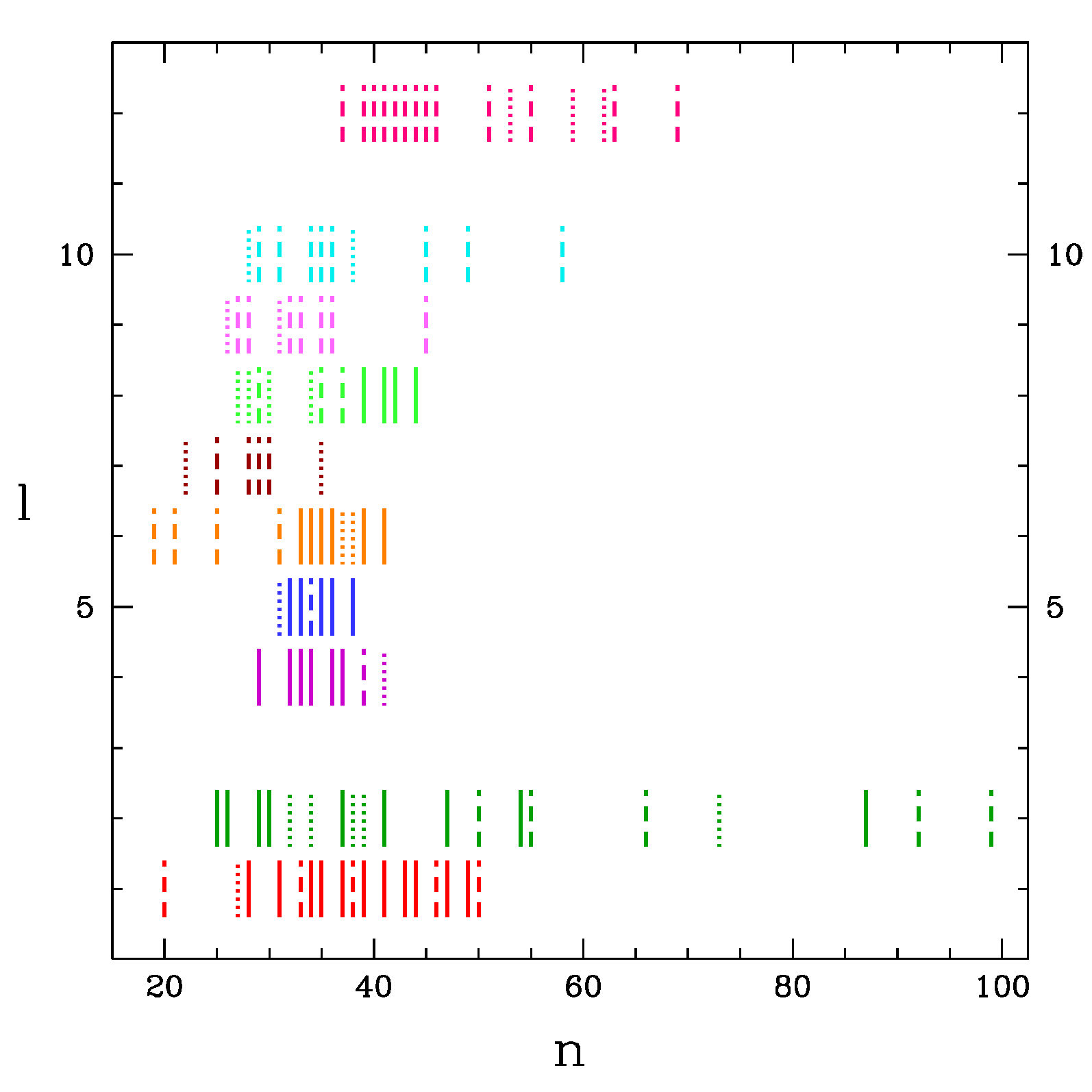}
\vspace{-4mm}
\caption{Summary of all identified g-modes. 
Dotted segments correspond to the modes with only 50\% confidence level 
(reported in brackets in Table~\ref{tab1}), dashed segments correspond to
the modes with a tentative identification.}
\label{fig6}
\end{figure}

When we plot the amplitude spectrum as a function of the pulsation period
(Fig.~\ref{fig2}, \ref{fig3} and \ref{fig4}), we clearly see sequences of
modes that are evenly spaced in period.
This corresponds to what is expected from theory in the asymptotic 
approximation ($n$$\gg$$l$) for high-order g-modes:
$\Delta$$P_l$=$\Delta$$\Pi$/$[l(l+1)]^{1/2}$, 
where $\Delta\Pi$ is the reduced period spacing, which is typically close to 
350~s for these stars.
In practice we can derive $\Delta$$\Pi$ by measuring $\Delta$$P_1$ or 
$\Delta$$P_2$ and then compute the expected $\Delta$$P_l$ for $l$$>$2.
From Fig.~\ref{fig2} we see that the two sequences with $l$=1 and $l$=2 are 
well defined, at least up to $\sim$12,500~s. 
Longer-period modes, up to $\sim$15,400~s, are present but their 
identification is less certain and it is not clear if these modes are
$l$=1 or $l$=2.
In both $l$=1 and $l$=2 sequences we do not see any clear signature of mode 
trapping in the high-order (long-period) region.
Looking at the \'echelle diagrams in Fig.~\ref{fig5} (upper plots),
only 3 modes appear to be shifted with respect to their normal position 
($l$=1, $n$=38,46 and $l$=2, $n$=99) but their identification is not certain.
As shown by \citet{charpinet14}, the lack of trapped modes in the high-order
region does not automatically mean that the star has a less stratified 
structure with respect to classical chemically stratified sdB models.
Lower-order $l$=1 or $l$=2 modes, with periods below 3,600~s, where mode
trapping could be more active, do not seem to be excited in this star.
The average period spacing that we obtain for $l$=1 and $l$=2, 256.5~s 
and 148.1~s, correspond to a basically identical reduced period spacing
of 362.75 and 362.77~s respectively, from which we can compute the expected
period spacing for the modes with higher degree.
In Fig.~\ref{fig3} the sequences with $l$=4, $l$=5, $l$=6 and $l$=8
are easily recognizable.
The $l$=7 sequence is visible at shorter periods in Fig.~\ref{fig4}.
The $l$=3 sequence is not seen.
%On the other hand we do not see the $l$=3 sequence, more sensitive to the 
%cancellation effects.
Considering that \kepler's response function extends from 4200 to 9000\,\AA\
with a maximum near 6000\,\AA, the fact that we do not see $l$=3 modes in our 
data is compatible with the expectation that these modes have much lower 
amplitudes in the optical at increasing wavelength \citep{randall05}.
However the same article predicts particularly high amplitudes in the red 
for the $l$=5 modes, which we do not see.
In Fig.~\ref{fig4} we see a large number of low-amplitude peaks at 
ever shorter periods, making difficult the mode identification in this region.
The density of modes implies that modes with high degree, up to at least 
$l$=12, are present in this star: we see sequences of at least 2 or 3 
consecutive peaks with $l$=7 and $l$=8 and partially also $l$=9, $l$=10 and 
$l$=12, while apparently we do not see consecutive $l$=11 modes.
More details are given in the caption of Fig.~\ref{fig4}.
In Fig.~\ref{fig5} the \'echelle diagrams of the sequences with 
$l$=1,2,4,5,6,7,8,9,10,12 are shown.
In Fig.~\ref{fig6} a summary of all the identified g-modes is given.
We note that for most degrees the excited modes have about the same range
of overtone index $n$.
This kind of properties can be useful for comparison with models in future
studies.
See for example Fig.~9, 11 and 12 of \citealt{jeffery06}, (although these 
authors consider only $l$=1,2,3,4 modes), or Fig.~6 of \citet{bloemen14}
from which, potentially, we might obtain also some indication on 
the evolutionary status of the star (although the stars considered in that 
article are much hotter than HD\,4539).

\subsection{Linear combinations}

In order to verify if some of the low-amplitude modes listed in 
Table~\ref{tab1} correspond to linear combinations of the main modes (those 
with an amplitude higher than 100 ppm), we computed f1+f2 and 2f1 for all 
the main modes.
%all the following combinations:
%f1+f2, 2f1, 2f1+f2, f1+2f2, 3f1, 3f1+f2, f1+3f2, 4f1.
When the difference (in absolute value) between a mode frequency and 
the linear combination is less than 0.15~$\mu$Hz (the formal frequency 
resolution), a flag is given in column 7 of Table~\ref{tab1}.
In particular, a linear combination may explain why we could not find an
identification for f133 and f138, the latter corresponding to three different
combinations.

\subsection{Frequency and amplitude time variations}

\begin{figure*}[ht]
\centering
%%%%%%%%%%%%%%%%%%%%%%%%%%%%%%%%%%%%%%%%%%%%%%%
\begin{minipage}[c]{50mm}
\vspace{-2mm}
\hspace{-15mm}
\includegraphics[width=45mm]{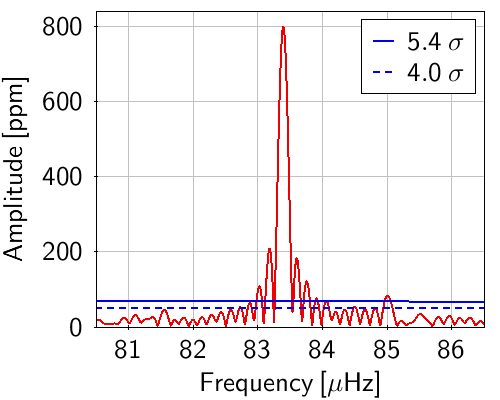}
\end{minipage}
\begin{minipage}[c]{50mm}
\vspace{-2mm}
\hspace{-5mm}
\includegraphics[width=45mm]{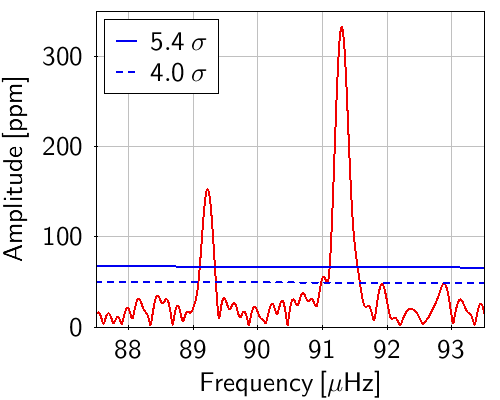}
\end{minipage}
\begin{minipage}[c]{50mm}
\vspace{-2mm}
\hspace{6mm}
\includegraphics[width=45mm]{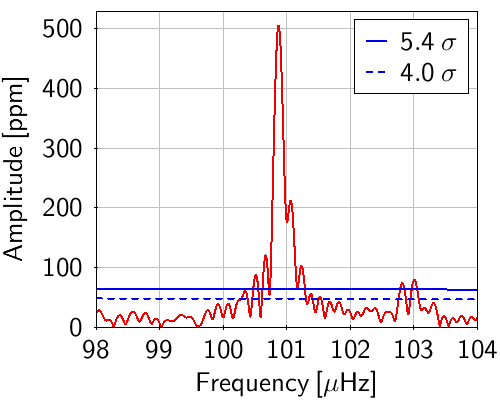}
\end{minipage}
%%%%%%%%%%%%%%%%%%%%%%%%%%%%%
\begin{minipage}[c]{50mm}
\vspace{-3.5mm}
\hspace{-11.2mm}
\includegraphics[width=50mm,angle=0]{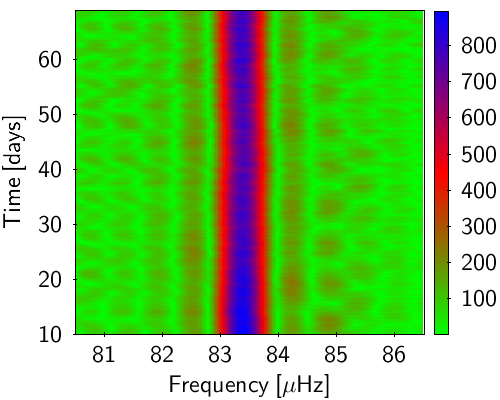}
\end{minipage}
\begin{minipage}[c]{50mm}
\vspace{-3.5mm}
\hspace{-1.2mm}
\includegraphics[width=50mm,angle=0]{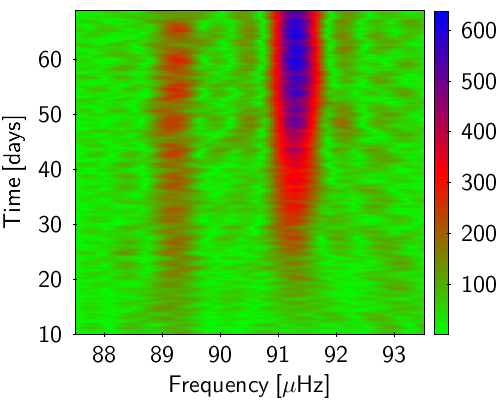}
\end{minipage}
\begin{minipage}[c]{50mm}
\vspace{-3.5mm}
\hspace{9.5mm}
\includegraphics[width=50mm,angle=0]{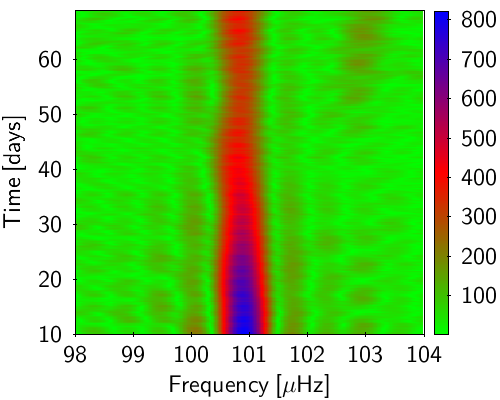}
\end{minipage}
%%%%%%%%%%%%%%%%%%%%%%%%%%%%%%%%%%%%%%%%%%%%%%%
\vspace{8mm}
%%%%%%%%%%%%%%%%%%%%%%%%%%%%%%%%%%%%%%%%%%%%%%%
\begin{minipage}[c]{50mm}
\vspace{3mm}
\hspace{-12.6mm}
\includegraphics[width=46mm]{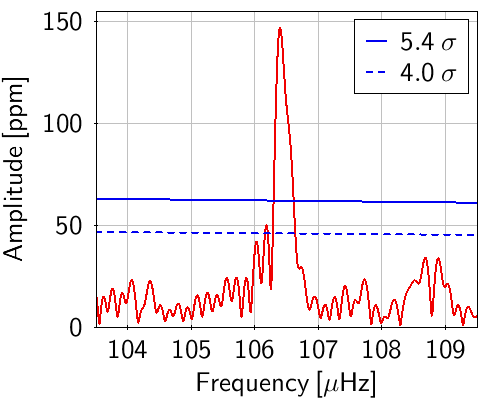}
\end{minipage}
\begin{minipage}[c]{50mm}
\vspace{3mm}
\hspace{-2.6mm}
\includegraphics[width=46mm]{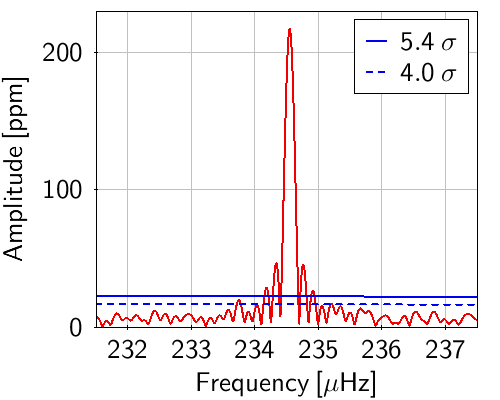}
\end{minipage}
\begin{minipage}[c]{50mm}
\vspace{3mm}
\hspace{8.1mm}
\includegraphics[width=46mm]{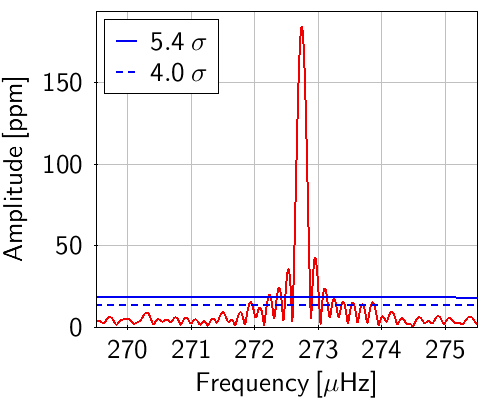}
\end{minipage}
%%%%%%%%%%%%%%%%%%%%%%%%%%%%%
\begin{minipage}[c]{50mm}
\vspace{-11.5mm}
\hspace{-11.25mm}
\includegraphics[width=50mm,angle=0]{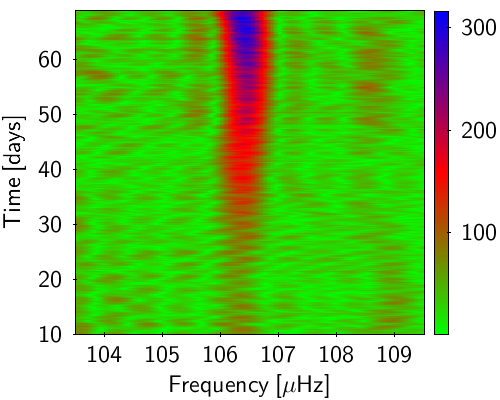}
\end{minipage}
\begin{minipage}[c]{50mm}
\vspace{-11.5mm}
\hspace{-1.2mm}
\includegraphics[width=50mm,angle=0]{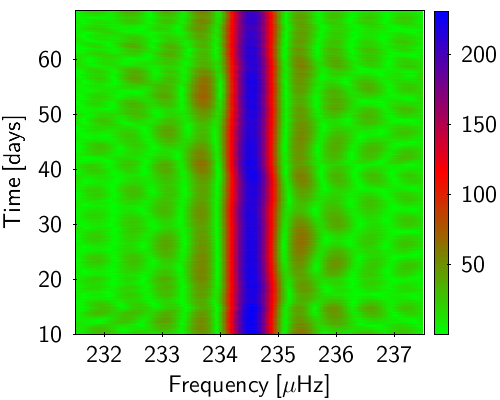}
\end{minipage}
\begin{minipage}[c]{50mm}
\vspace{-11.5mm}
\hspace{9.4mm}
\includegraphics[width=50mm,angle=0]{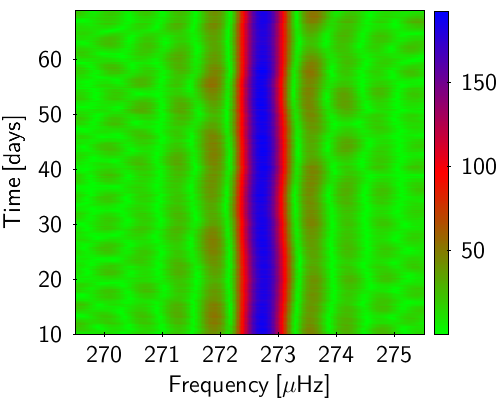}
\end{minipage}
%%%%%%%%%%%%%%%%%%%%%%%%%%%%%%%%%%%%%%%%%%%%%%%
\vspace{8mm}
%%%%%%%%%%%%%%%%%%%%%%%%%%%%%%%%%%%%%%%%%%%%%%%
\begin{minipage}[c]{50mm}
\vspace{3mm}
\hspace{-12.65mm}
\includegraphics[width=46mm]{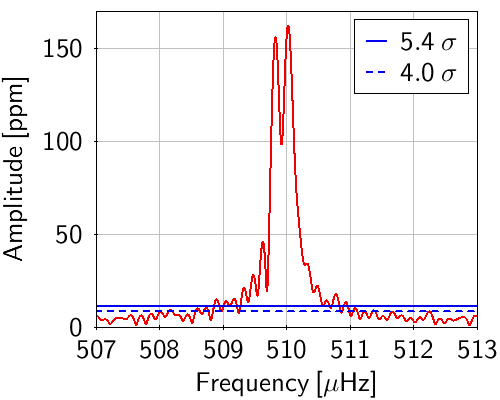}
\end{minipage}
\begin{minipage}[c]{50mm}
\vspace{3mm}
\hspace{-1.53mm}
\includegraphics[width=46mm]{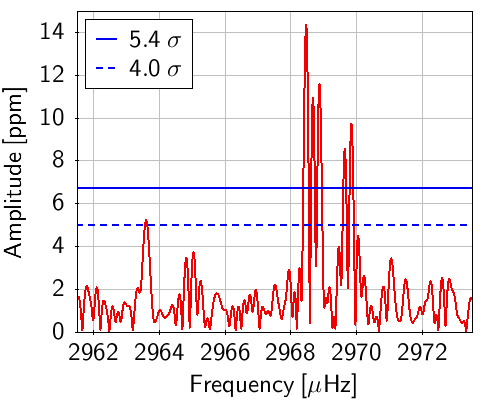}
\end{minipage}
\begin{minipage}[c]{50mm}
\vspace{3mm}
\hspace{9.1mm}
\includegraphics[width=46mm]{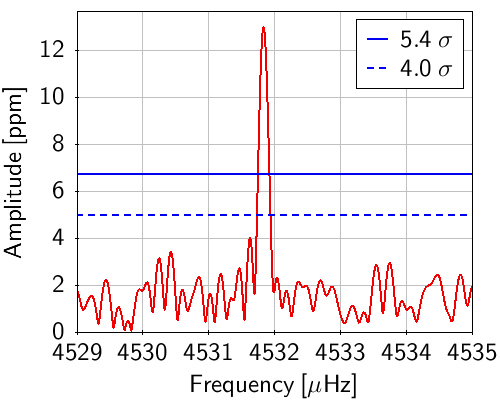}
\end{minipage}
%%%%%%%%%%%%%%%%%%%%%%%%%%%%%
\begin{minipage}[c]{50mm}
\vspace{-11.5mm}
\hspace{-10.8mm}
\includegraphics[width=50mm,angle=0]{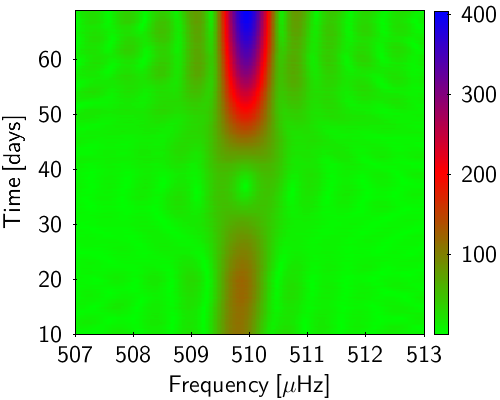}
\end{minipage}
\begin{minipage}[c]{50mm}
\vspace{-11.5mm}
\hspace{-1.2mm}
\includegraphics[width=50mm,angle=0]{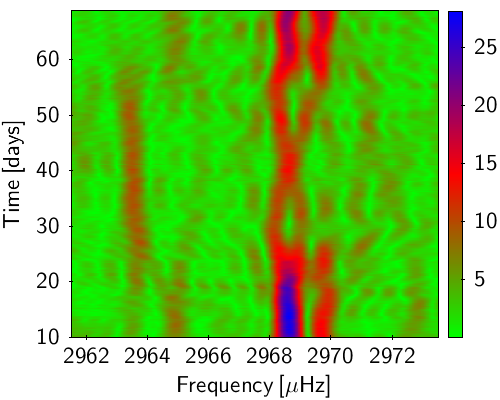}
\end{minipage}
\begin{minipage}[c]{50mm}
\vspace{-11.5mm}
\hspace{9.4mm}
\includegraphics[width=50mm,angle=0]{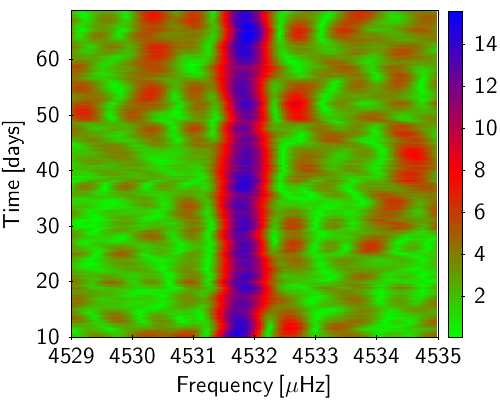}
\end{minipage}
%%%%%%%%%%%%%%%%%%%%%%%%%%%%%%%%%%%%%%%%%%%%%%%
\vspace{3mm}
%%%%%%%%%%%%%%%%%%%%%%%%%%%%%%%%%%%%%%%%%%%%%%%
\caption{Sliding FTs of some of the main pulsation modes. From from top left 
to bottom right f163, f160, f157 and f154 ($l$=1); f132 and f130 ($l$=2); 
f109 ($l$=6); f6-f11 and f1 (p-modes). 
We see that some modes, like f130, f132 and partially f163 are stable both 
in frequency and amplitude, while others show variations that are 
particularly strong in amplitude.
The color coded amplitude is given in ppm.
%the highest-amplitude signal (f163) at 83.396\muHz.
}
\label{fig7}
\end{figure*}

From the continuous light curves produced by $Kepler$ and $K2$ we have 
learned that oscillation frequencies in sdB stars are not as stable as 
previously believed (see e.g. \citealt{reed14}) and, at least in one case,
they may even show a stochastic behavior \citep{ostensen14}.
A systematic study of these aspects is presented by \citet{zong18}.
Even though the much longer $Kepler$ light curves are more suitable
to study these aspects, a short analysis of the frequency/amplitude
variations of HD\,4539 was performed using sliding Fourier 
Transforms (sliding FTs) to see how the pulsation frequencies and amplitudes 
vary over the course of the 78.7 days of observation. 
We divided the $K2$ light curve into 30 subsets of about 20.7 days each,
moving forward the beginning of each subset of two days at each step,
and we computed the amplitude spectrum of each subset.
From Table~\ref{tab1} we selected a few high-amplitude frequencies 
which are particularly variable (those marked with "SR=Strong Residuals after 
prewhitening" in the last column of Table~\ref{tab1}) or, on the contrary, 
particularly stable (marked with ``NoR=No Residuals'' in the last column of 
Table~\ref{tab1}).
The sliding FTs of these frequencies are shown in Fig.~\ref{fig7}.

\section{Low-resolution spectroscopy, SED, and atmospheric parameters} 

\begin{figure}
\centering
\includegraphics[width=8.6cm,angle=0]{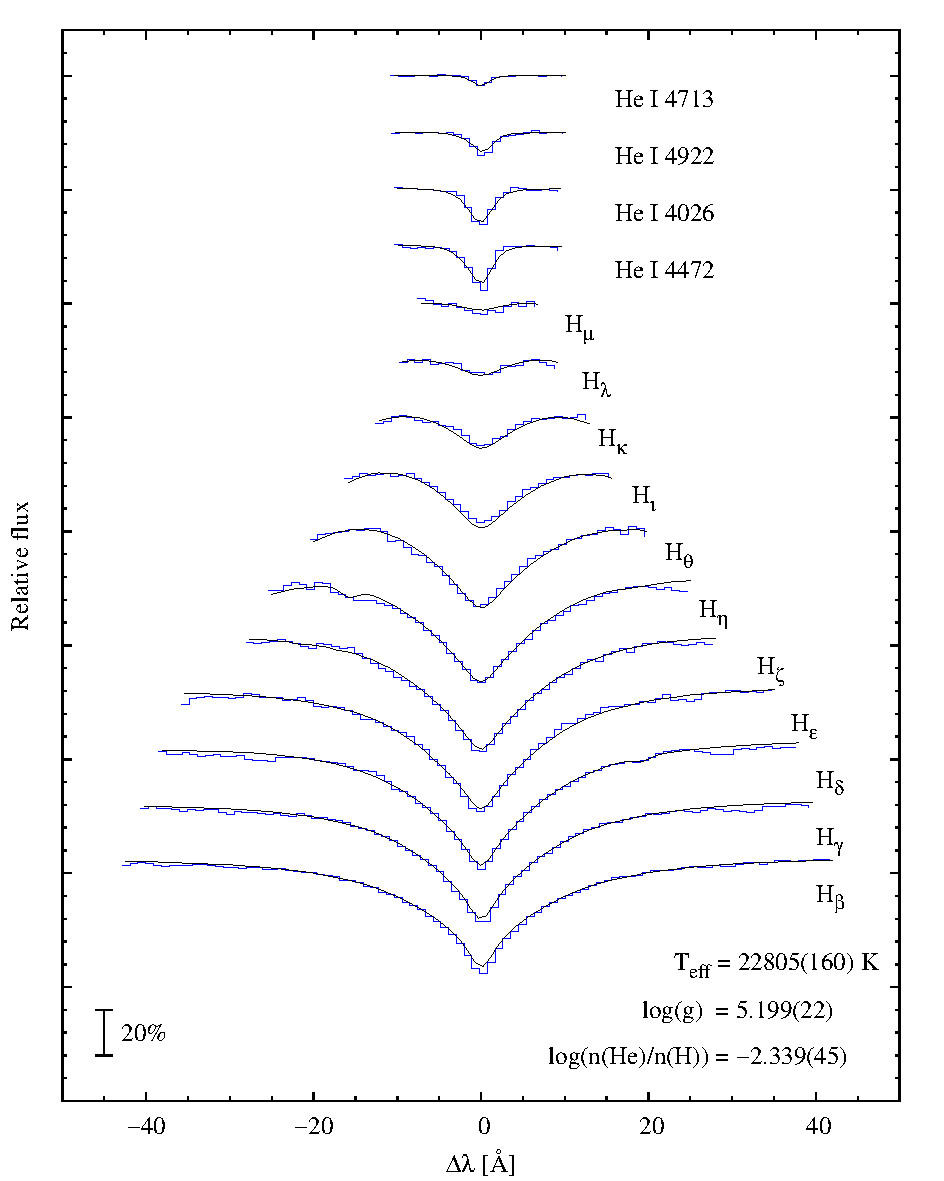}
\vspace{-4mm}
\caption{Atmospheric parameters of HD\,4539 as obtained from the sum
of 3 ALFOSC spectra.}
\label{fig_lrs}
\end{figure}

\begin{figure*}
\centering
\includegraphics[width=17.2cm,angle=0]{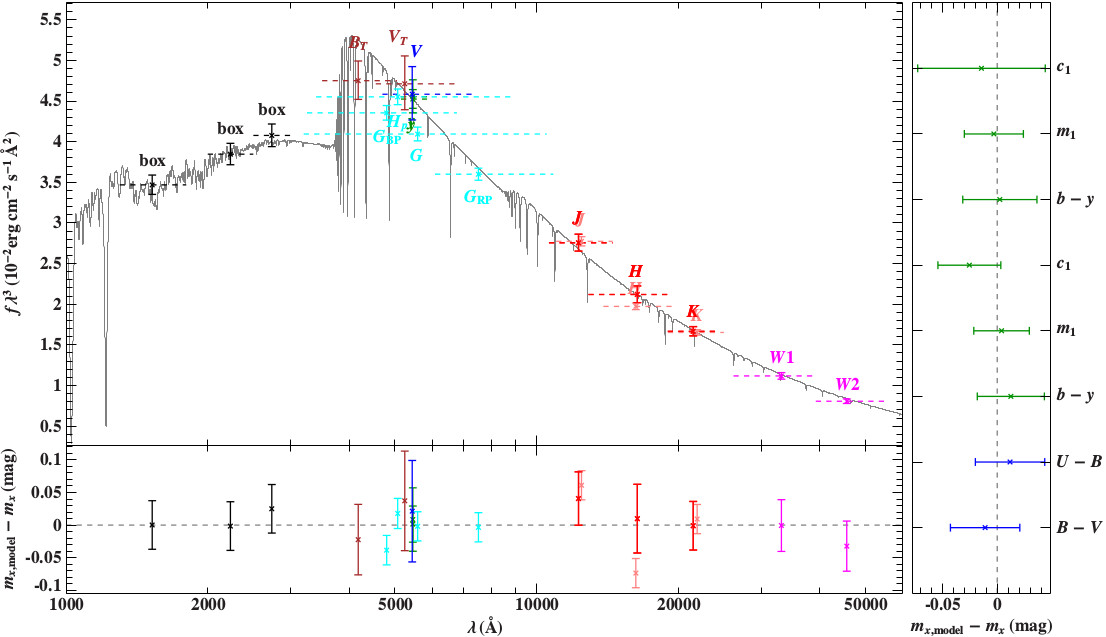}
%\vspace{-4mm}
\caption{Spectral energy distribution of HD~4539. 
Colored data points represent the filter-averaged fluxes, which were converted 
from observed magnitudes (the filter widths are indicated by dashed horizontal 
lines), while the gray solid line represents a synthetic spectrum computed from
a model atmosphere. 
The 3 box ``filters'' in the UV band, which cover the range
1300-1800, 2000-2500, and 2500--3000~\AA, are extracted from IUE spectra
(see \citealt{heber18} for more details).
The bottom/right panels show the differences between synthetic and observed 
magnitudes/colors. 
The following color codes are used to identify the photometric systems: 
UV (black), Tycho (brown), Johnson (blue), Hypparcos/Gaia (cyan), Str\"{o}mgren
(green), 2MASS (red), and WISE (magenta).}
\label{fig_sed}
\end{figure*}

Given its brightness and long history in sdB literature, there are several 
determinations of the atmospheric parameters of HD\,4539, sometimes with 
significant differences:
\teff=25,000$\pm$2000~K, \logg=5.4$\pm$0.2 \citep{baschek72};
\teff=24,800~K, \logg=5.4, log($N$(He)/$N$(H))=--2.32$\pm$0.05 
\citep{heber86};
\teff$\simeq$27,000~K, \logg$\simeq$5.46, log($N$(He)/$N$(H))=--2.30 
\citep{saffer94};
\teff=25,200~K, \logg=5.40 \citep{cenarro07};
\teff=26,000$\pm$500~K, \logg=5.2$\pm$0.1,  
log($N$(He)/$N$(H))=--2.32$\pm$0.05 \citep{sale08};
%(equal to [He/H]=--1.25$\pm$0.05) 
%
% Caroline Pereira's thesis % NOT FOUND !! -> ASK Simon !!
\teff=23,000~K \citep{geier12};
\teff=23,200$\pm$100~K, \logg=5.20$\pm$0.01,
log($N$(He)/$N$(H))=--2.27$\pm$0.24 \citep{schneider18}.

As part of our K2 sdBV follow-up spectroscopic survey \citep{telting14b}, 
we did a new determination of the atmospheric parameters of HD~4539 using 
three low-resolution spectra (R$\sim$2000, or 2.2\,\AA) taken at the 2.56\,m 
Nordic Optical Telescope (NOT, La Palma) with ALFOSC, grism\#18, 0.5 arcsec 
slit, and CCD\#14, giving an approximate wavelength range 345-535 nm.  
The observations were carried out on the night starting on 2016-12-07.
The spectra were homogeneously reduced and analysed. Standard
reduction steps within IRAF include bias subtraction, removal of
pixel-to-pixel sensitivity variations, optimal spectral extraction,
and wavelength calibration based on helium arc-lamp spectra. 
The peak signal-to-noise ratio of the individual spectra is in excess of 200.
By fitting 11 Balmer lines and 4 He\,{\sc I} lines (Fig.~\ref{fig_lrs}) we 
obtain \teff=22,800$\pm$160~K, \logg=5.20$\pm$0.02 and 
log($N$(He)/$N$(H))=--2.34$\pm$0.05, in good agreement with 
\citet{schneider18}.
%
%By fitting 11 Balmer lines and 4 He\,{\sc I} lines (Fig.~\ref{fig_lrs}) we 
%obtain $T_{\mathrm eff}\,=\,22,800$\pm$160 K, log$g$\,=\,5.20$\pm$0.02 and 
%log($N$(He)/$N$(H))\,=\,$-$2.34$\pm$0.05. 
%
The values that we obtain are relative to the H/He LTE grid of 
\citet{heber00}.
The errors are the formal fitting errors, which only reflect the S/N of the 
mean spectrum and the match to the model, and not any systematic effects 
caused by the assumptions underlying those models, which can be an order of 
magnitude larger.

An independent determination of \teff\ was derived also by fitting, 
with an appropriate model atmosphere, the spectral energy distribution (SED) 
obtained from available photometric measurements covering all wavelenghts 
from the ultraviolet to the infrared (Fig.~\ref{fig_sed}), following the method
described by \citealt{heber18} (see also \citealt{schindewolf18}).
Here we just mention the sources of the photometric data.
We used ultraviolet fluxes extracted from IUE spectra (downloaded from the 
``Mikulski Archive for Space Telescopes'' (MAST)), Tycho $B_T$, $V_T$ 
magnitudes and Hypparcos $H_P$ magnitude \citep{esa97}, Gaia DR2 magnitudes 
($G$, $G_{BP}$, $G_{RP}$, see \citealt{gaia18}), 2MASS J, H, K 
(\citealt{skrutskie06}) and WISE W1, W2 \citep{cutri13} infrared magnitudes.
Johnson V magnitude and colors and Str\"{o}mgren colors from VizieR 
\citep{ochsenbein00} were also included. 
Adopting \logg=5.20$\pm$0.05, we obtain \teff=23,470$^{+650}_{-210}$~K,
an angular diameter 
$\Theta$=$\left(6.38^{+0.06}_{-0.10}\right)$$\times$10$^{-11}$~rad 
and zero interstellar reddening (E(B--V)$<$0.009).
If we combine the angular diameter with the Gaia DR2 
%distance of 185.7$\pm$4.6~pc
parallax $\varpi$=5.384$\pm$0.132\,mas 
%(or d=185.7$\pm$4.6~pc),
(or d=185.7$^{+4.7}_{-4.4}$~pc),
we obtain for HD~4539 a radius of 0.263$^{+0.009}_{-0.011}$~\rsun\ and a mass 
of 0.40$\pm$0.08~\msun\ from the relation 
%M=g\,$\Theta^2$/(4\,G\,$\varpi^2$).
M=$g$\,$\Theta^2$d$^2$/(4\,G).
 
\section{High-resolution spectroscopy and radial velocities}

\begin{figure}
\centering
\includegraphics[width=8.45cm,angle=0]{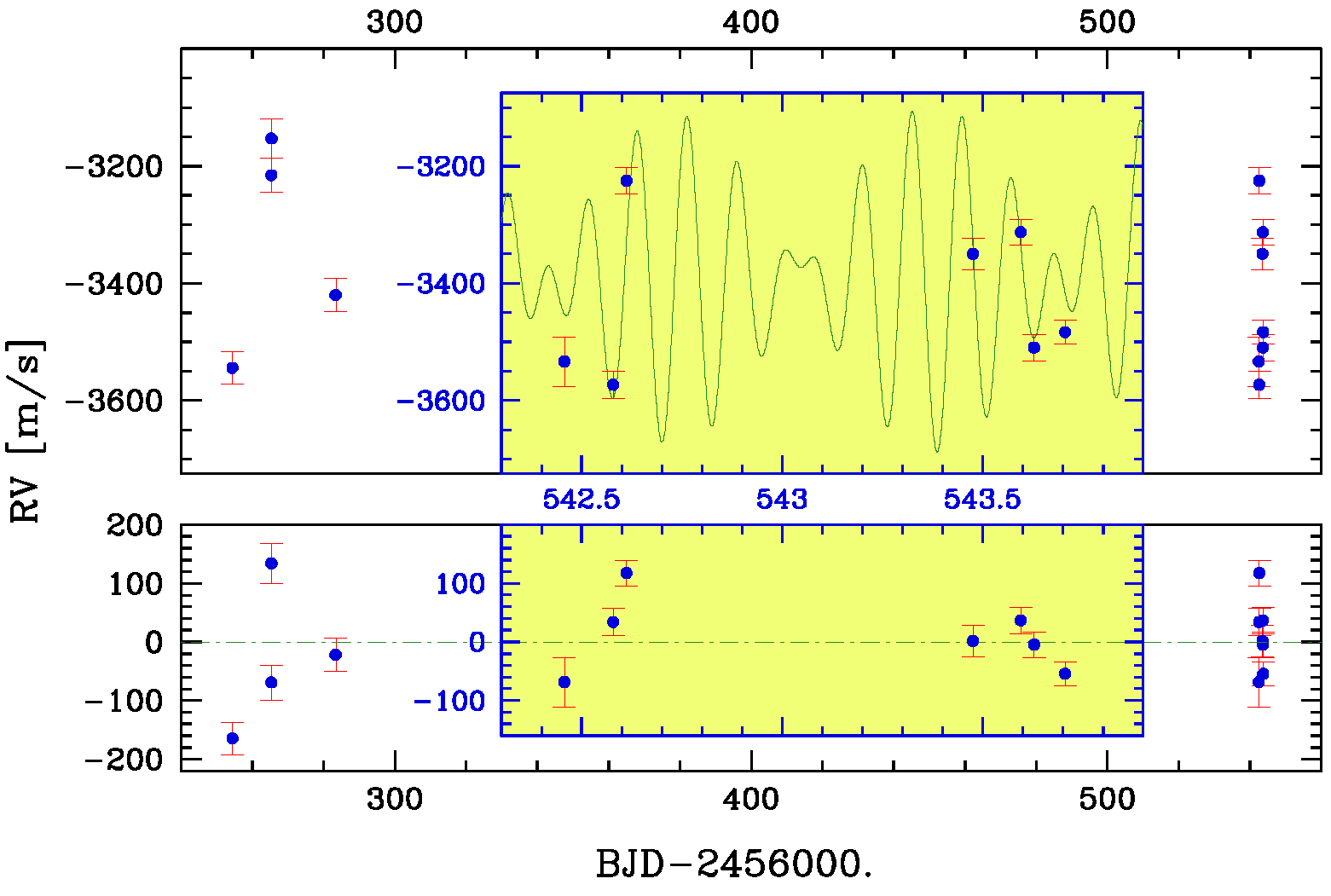}
\vspace{-4mm}
\caption{Upper panel: radial velocities of HD\,4539 and best sinusoidal fit
using the 3 highest-amplitude pulsation frequencies from Table~\ref{tab1} 
(f163, f157 and f158).
Lower panel: residuals.
The 2 yellow insets show a magnification of the 2nd run.}
\label{fig_rvs}
\end{figure}

\begin{figure}
\centering
\includegraphics[width=8.6cm,angle=0]{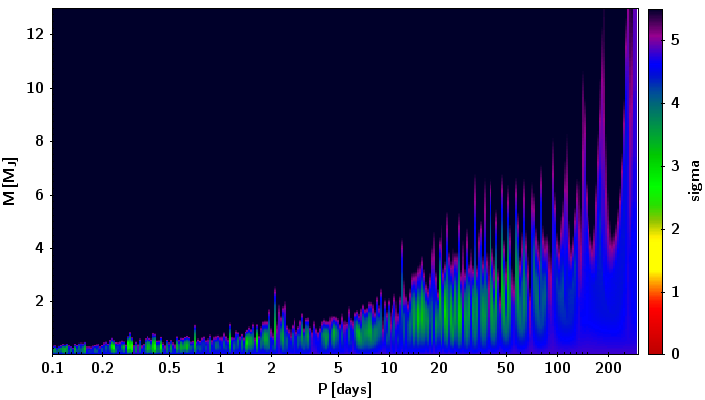}
\includegraphics[width=8.6cm,angle=0]{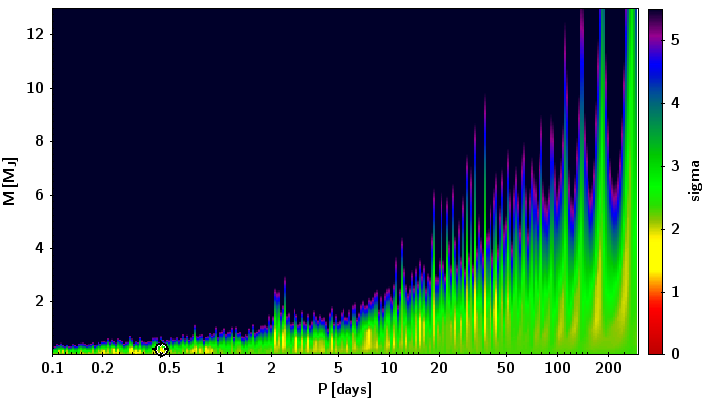}
\vspace{-4mm}
\caption{Upper limits to the mass of a hypothetical companion to 
HD\,4539 as a function of orbital period.
The upper panel is obtained using the RV data (upper panel of 
Fig.~\ref{fig_rvs}) while in the lower panel we used the RV residuals
(lower panel of Fig.~\ref{fig_rvs}).
The regions where the presence of a companion would be best compatible with 
the RV measurements are those in yellow/green.
The counterintuitive fact that the ``good'' regions are smaller in the upper 
panel highlights the difficulty of fitting the RV data with a single orbital 
frequency when the RV variations are caused by several pulsation frequencies.
In the lower panel a viewfinder at P$\simeq$0.45~d and 
M\,sin$i$$\simeq$0.15~\mjup\ shows the best point in terms of minimum 
difference between RV residual and synthetic RV value. See text for more 
details.}
\label{fig_comp}
\end{figure}

HD\,4539 was observed with Harps-N at the Telescopio Nazionale Galileo 
(TNG, La Palma) in 2 runs (November-December 2012 and September 2013) and we 
collected in total 11 high-resolution spectra with a mean signal-to-noise 
ratio of 102.
While a chemical abundance analysis is beyond the scope of this article,
here we concentrate on the radial velocity (RV) measurements.
Using the cross correlation function on more than 200 absorption lines
(excluding H and He lines that are too broad), we computed the RVs of the 
star and we found a system velocity of --3392.7 m/s with significant variations
around this value that are attributed to the g-mode pulsations 
(Fig.~\ref{fig_rvs}).
With only 11 points it is obviously not possible to determine the RV 
amplitude of each pulsation mode.
However, by fitting these 11 RVs with the three highest-amplitude pulsation 
frequencies from Table~\ref{tab1}, we can at least obtain a zero-order 
measurement of the RV amplitudes involved.
Assuming constant frequencies, we obtain amplitudes of 146, 164 and 40 m/s 
for f163, f157 and f158 respectively.

The residuals shown in the lower panel of Fig.~\ref{fig_rvs} can be used 
to obtain an upper limit to the minimum mass (M\,sin$i$) of a hypothetical 
companion.
We computed a series of synthetic RV curves for different orbital periods
and companion masses and compared these curves with the RV residuals.
For each synthetic RV curve we selected the best phase using a weighted least 
squares algorithm. For each observational point we computed the difference, 
in absolute value and in $\sigma$ units (where $\sigma$ is the observation 
error), between RV residual and synthetic RV value. The color coding in 
Fig.~\ref{fig_comp} corresponds to the mean value of this difference.
We should keep in mind, however, that these upper limits to the mass of a 
companion are likely overestimated given that the residuals shown in the
lower panel of Fig.~\ref{fig_rvs} still contain some residual signal due to 
the many pulsation modes that were not considered in our fit.

\section{Summary}

The analysis of the $K2$ data on HD\,4539 shows a very rich spectrum, 
with several g-modes of high-degree, up to at least $l$=12.
To our knowledge, this is the first time that $l$=12 modes are seen 
in an sdB star and that sequences of consecutive modes with $l\ge$4, up to 
at least $l$=8, are clearly recognized.
G-modes with such high degree are very rare also in other types of pulsating 
stars except, perhaps, $\delta$ Scuti stars (see e.g. 
\citealt{dziembowski98}; \citealt{mantegazza12}).
The identification of these high-degree g-modes in HD\,4539 is made possible 
by the absence of rotational splitting of the frequencies, which makes the 
spectrum cleaner, and by the star brightness, which makes it possible to detect
amplitudes below 10 ppm.
Therefore this star represents a challenge and an ideal laboratory to test
current asteroseismic models.
Thanks to the simultaneous presence of a few p-modes as well,
potentially both the internal layers near the core and the external layers
of the star can be probed through seismic tools.
%The absence of trapped modes is an aspect that may deserve further study.
We have not been able to identify any trapped modes in the sequences of 
HD\,4539, but such identifications are extremely hard when the mode sequences 
are incomplete and the order of individual modes cannot be verified with 
rotational splittings.

The absence of multiplets points towards a rotation period longer than the
$K2$ light curve and/or a very low inclination.
This is further strengthened by the presence of a few high-amplitude peaks in 
the Fourier spectrum that are completely removed when subtracting a single 
sinusoidal wave, suggesting that there are no unresolved multiplets 
in these cases.
When we consider the rotation velocity obtained by \citet{geier12}, 
if the line broadening they measured was actually due only to rotation it 
would imply an extremely low inclination and a fast rotation with a period 
of the order of hours, which appears unlikely, also considering that part of 
the line broadening is produced by the pulsations.

Our new determination of the atmospheric parameters of HD\,4539 confirms that
this star is close to the low-temperature boundary of the g-mode instability 
strip, where the p-modes are normally not present. We know only one other sdB 
pulsator, KIC~2697388 (alias SDSS~J190907.14+375614.2), showing both g- and 
p-modes at \teff\ below $\sim$24,000~K \citep{kern17}.
The reason why only these 2 relatively cool stars show p-modes is not clear.
In the case of HD\,4539 its brightness is certainly helpful in detecting 
ppm-level modes, but to answer this question we probably need larger statistics
and the $TESS$ mission can help in this respect.

The RV measurements obtained from high-resolution spectroscopy show 
significant variations due to the pulsations, with amplitudes of the order
of 150~m/s for the main modes, and allow us to exclude the presence of a 
companion with a minimum mass higher than a few Jupiter masses for orbital 
periods below $\sim$300 days.

\section*{Acknowledgements}

The $K2$ data presented in this paper were obtained from the Mikulski Archive 
for Space Telescopes (MAST). Space Telescope Science Institute is operated by 
the Association of Universities for Research in Astronomy, Inc., under NASA 
contract NAS5-26555.
The spectroscopic results are based on observations collected 
at the Telescopio Nazionale Galileo (TNG, AOT26/TAC41 and AOT28/TAC23), 
operated by the Centro Galileo Galilei of the Istituto Nazionale di 
Astrofisica (INAF), and at the Nordic Optical Telescope (NOT), operated 
jointly by Denmark, Finland, Iceland, Norway, and Sweden.
Both the TNG and the NOT are operated on the island of La Palma 
at the Spanish Observatorio del Roque de los Muchachos (ORM)
of the Instituto de Astrofísica de Canarias (IAC).
ASB gratefully acknowledges financial support from the Polish National Science
Center under projects No.\,UMO-2017/26/E/ST9/00703 and UMO-2017/25/B 
ST9/02218.

%%%%%%%%%%%%%%%%%%%%%%%%%%%%%%%%%%%%%%%%%%%%%%%%%%

%%%%%%%%%%%%%%%%%%%% REFERENCES %%%%%%%%%%%%%%%%%%

% The best way to enter references is to use BibTeX:

%\bibliographystyle{mnras}
%\bibliography{sdBrefs} % if your bibtex file is called sdBrefs.bib

% Alternatively you could enter them by hand, like this:
% This method is tedious and prone to error if you have lots of references

%%%%%%%%%%%%%%%%%%%%%%%%%%%%%%%%%%%%%%%%%%%%%%%%%%

%%%%%%%%%%%%%%%%% APPENDICES %%%%%%%%%%%%%%%%%%%%%

\appendix

\section{Some extra material}

%%%%%%%%%%%%%%%%%%%%%%%%%%%%%%%%%%%%%%%%%%%%%%%%%%

% Don't change these lines
\bsp	% typesetting comment
\label{lastpage}
\end{document}

%% file: defs.tex
\newcommand{\kepler}{{\em Kepler}}

\newcommand{\teff}{\ensuremath{T_{\mathrm{eff}}}}
\newcommand{\logg}{\ensuremath{\log g}}

\newcommand{\lsim}{\raisebox{-1ex}{$\stackrel{{\displaystyle<}}{\sim}$}}

\newcommand{\msun}{${\mathrm{M}}_{\odot}$} 
\newcommand{\rsun}{${\mathrm{R}}_{\odot}$} 
 
\newcommand{\muHz}{$\mu$Hz}

\newcommand{\mjup}{$\rm M_{\rm Jup}$}

\def\mnras{MNRAS}% Monthly Notices of the RAS
\def\aj{AJ}%
\def\apj{ApJ}%
\def\apjl{ApJ}%
\def\aap{A\&A}%
\def\pasp{PASP}%